\documentclass[acmsmall]{acmart}
\usepackage{booktabs} 
\usepackage{geometry}
\usepackage{textcomp} 
\usepackage[table]{xcolor}

\newcommand{\add}[1]{\textcolor{black}{#1}}
\newcommand{\newadd}[1]{\textcolor{black}{#1}}

\AtBeginDocument{%
  }

\begin{document}

\title[How Multi-Channel Networks (MCNs) Govern Live-Streaming Labor in China]{\newadd{Constructing Algorithmic Authority: How Multi-Channel Networks (MCNs) Govern Live-Streaming Labor in China}}

\author{Qing Xiao}
\affiliation{
  \institution{Human-Computer Interaction Institute, Carnegie Mellon University}
  \city{Pittsburgh}
  \state{Pennsylvania}
  \country{USA}
}
\email{qingx@cs.cmu.edu}

\author{Rongyi Chen}
\thanks{This work was completed during Rongyi Chen’s visiting research at George Mason University.}
\affiliation{
  \institution{Central South University}
  \city{Changsha}
  \country{China}
}
\email{chenrongyiwork@gmail.com}

\author{Jingjia Xiao}
\affiliation{
  \institution{Department of Sociology, University of California San Diego}
  \city{La Jolla}
  \state{California}
  \country{USA}
}
\email{jix082@ucsd.edu}

\author{Tianyang Fu}
\affiliation{
  \institution{Moody College of Communication, The University of Texas at Austin}
  \city{Austin}
  \state{Texas}
  \country{USA}
}
\email{tianyangfu@utexas.edu}

\author{Alice Qian}
\affiliation{
  \institution{Human-Computer Interaction Institute, Carnegie Mellon University}
  \city{Pittsburgh}
  \state{Pennsylvania}
  \country{USA}
}
\email{aqzhang@andrew.cmu.edu}

\author{Xianzhe Fan}
\affiliation{
  \institution{The University of Hong Kong}
  \thanks{This work was completed during Xianzhe Fan’s visiting research at Carnegie Mellon University.}
  \country{Hong Kong SAR, China}
}
\email{1038716025@qq.com}

\author{Bingbing Zhang}
\affiliation{
  \institution{School of Journalism and Mass Communication, University of Iowa}
  \city{Iowa City}
  \state{Iowa}
  \country{USA}
}
\email{bingbing-zhang@uiowa.edu}

\author{Zhicong Lu}
\affiliation{
  \institution{Department of Computer Science, George Mason University}
  \city{Fairfax}
  \state{Virginia}
  \country{USA}
}
\email{zlu6@gmu.edu}

\author{Hong Shen}
\affiliation{
  \institution{Human-Computer Interaction Institute, Carnegie Mellon University}
  \city{Pittsburgh}
  \state{Pennsylvania}
  \country{USA}
}
\email{hongs@cs.cmu.edu}

\renewcommand{\shortauthors}{Xiao et al.}

\begin{abstract}
\newadd{This study examines the discursive construction of algorithms and its role in labor management in Chinese live-streaming industry by focusing on how intermediary organizations (Multi-Channel Networks, MCNs) actively construct, stabilize, and deploy particular interpretations of platform algorithms as instruments of labor management. Drawing on a nine-month ethnographic fieldwork and 44 interviews with live-streamers, former live-streamers, and MCN staff, we examine how MCNs produce and circulate structured interpretations of platform algorithms across organizational settings. We show that MCNs articulate two asymmetric yet interconnected forms of algorithmic interpretations. Internally, MCNs managers approach algorithms as volatile and uncertain systems and adopt probabilistic strategies to manage performance and risk. Externally, in interactions with streamers, MCNs circulate simplified and prescriptive algorithmic narratives that frame platform systems as transparent, fair, and responsive to individual effort. These organizationally produced algorithmic interpretations are embedded into training materials, live-streaming performance metrics, and everyday management practices. Through these mechanisms, streamers internalize responsibility for outcomes, intensify self-discipline, and increase investments in equipment, performing skills, and routines to maintain streamer-audience relationship, while accountability for unpredictable outcomes is increasingly shifted away from managers and platforms. This study contributes to CSCW and platform labor research by demonstrating how discursively constructed algorithmic knowledge can function as an intermediary infrastructure of soft control, shaping how platform labor is regulated, moralized, and governed in practice.}
\end{abstract}



\maketitle

\section{Introduction}

\newadd{As algorithmic systems have increasingly become central to digital labor, workers across domains such as ride-hailing, e-commerce, and live-streaming are required to operate within opaque, volatile, and hard-to-interpret systems of performance and evaluation. Algorithms embed managerial logic into platform infrastructures by shaping what becomes visible, how performance is assessed, and which forms of labor receive recognition or reward \cite{wood2019good, bucher2018if, kellogg2020algorithms}. For workers, everyday labor unfolds under conditions of uncertainty, where outcomes appear closely tied to shifting algorithmic signals. These conditions give rise to sustained interpretive work as workers seek to make sense of how algorithmic systems operate and how their activity may be evaluated.}

\newadd{Scholars in HCI, CSCW, and digital studies have documented how workers develop practical understandings of algorithmic systems through experience. Across this literature, these interpretive processes have been theorized using concepts such as algorithmic imaginaries \cite{bucher2018if}, mental models \cite{ngo2020exploring}, and folk theories of algorithms \cite{devito2017algorithms}. Empirically, this body of research shows how workers exchange interpretations within peer networks \cite{mcdaid2023algorithmic, yao2021together}, use shared experiences and explanations to rationalize algorithmic outcomes \cite{christin2020metrics, poon2021computer}, and align their practices with perceived algorithmic preferences \cite{bonini2024algorithms, lei2021delivering}. Recent work further highlights the collective nature of these interpretive processes, demonstrating how understandings of algorithms circulate within communities and become stabilized through repeated interaction \cite{xiao2024let, zhang2023push}. Together, these studies establish algorithmic interpretation as a central component of contemporary platform labor.} 

\newadd{Moreover, prior work also treats algorithms as discursively constructed objects, examining how institutions articulate and frame particular understandings of algorithmic systems through their explanations \cite{ananny2018seeing, cellard2022algorithms, schulz2023new}. }\newadd{\citeauthor{lin2025vernacular} (\citeyear{lin2025vernacular}) introduces the term “discursive construction of algorithms” to capture the discursively produced character of algorithmic interpretations. }\newadd{\citeauthor{lin2025vernacular} (\citeyear{lin2025vernacular}) shows that Douyin’s official accounts foreground algorithmic fairness and effort-responsiveness while downplaying uncertainty, thereby reducing users’ suspicion of unfairness and aligning user behavior with the platform’s profit model.} \newadd{While this perspective clarifies how platforms shape users’ orientations toward algorithmic systems, we still know much less about how such discursive constructions become integrated into labor management through intermediary organizations.} 

\newadd{Within platformized labor markets, intermediary organizations often sits between platform-facing discourse with everyday work practices. Through training programs, guidelines, and managerial communication, these organizations curate algorithmic explanations and define credible knowledge for workers. These dynamics foreground questions about how interpretive authority of algorithms is organized and how discursively constructed algorithmic knowledge becomes parts of labor management within institutions.} 

This paper addresses this gap by examining the role of Multi-Channel Networks (MCNs) in China’s live-streaming industry. MCNs are intermediary organizations that operate between individual streamers and platform companies. They help streamers initiate their channel, improve their content, build their personal brand, and make profits, while also shaping what streamers produce and how they work \cite{zhang2024intermediated,zhang2024contesting,liang2024manufacturing,wang2024harnessing}. While MCNs have been widely studied in western contexts such as YouTube as service providers \cite{siciliano2023intermediaries,gardner2016whats}, less attention has been paid to 
their epistemic role in algorithm interpretation. We advance this literature by theorizing MCNs not just as economic intermediaries, but as organizational actors that actively interpret and institutionalize algorithmic knowledge. This shift foregrounds how algorithmic interpretation is produced and circulated at scale. China’s live-streaming industry, then, offers a particularly generative setting for this analysis. Live-streaming in China is large scale, highly industrialized, and labor-intensive \cite{ye2023navigating,sun2023unraveling}, with MCNs exerting strong control over labor and visibility. Despite their waning influence on Western markets \cite{alexander2018youtube}, MCNs remain influential in China, shaping how streamers understand and respond to algorithmic systems. \newadd{Given the informational asymmetries in live-streaming work and the limited access many streamers have to reliable explanations of platform algorithms -- including
many women from rural areas with limited formal education \cite{xiao2024happens,tang2022dare}, MCNs can possess disproportionate epistemic authority in shaping what counts as credible algorithmic knowledge.}

\newadd{Building on such research potential, this study investigates how Chinese MCNs discursively construct algorithms and integrate these constructions into organizational practices in labor management. Drawing on nine months of ethnographic fieldwork and 44 interviews with live-streamers, former live-streamers, and MCN staff, we ask:}
\begin{itemize}
\item \newadd{\textbf{RQ1:} How do MCNs construct organizational interpretations of algorithms?}
\item \newadd{\textbf{RQ2:} How are these interpretations embedded into everyday labor management practices?}
\item \newadd{\textbf{RQ3:} How do live-streamers react to MCNs’ algorithmic interpretations?}
\end{itemize}

\newadd{In response to \textbf{\textit{RQ1}}, We found that MCNs construct organizational interpretations of algorithms through a dual framework. Internally, MCN staff treat algorithms as volatile and uncertain systems, relying on probabilistic reasoning to manage performance and organizational risk. Externally, in their interactions with cooperated streamers, MCNs promote simplified and prescriptive narratives that portray platform algorithms as transparent, fair, and responsive to individual effort.}

\newadd{In response to \textbf{\textit{RQ2}}, we demonstrate how these internal and external algorithmic interpretations are embedded into everyday labor management practices. MCNs translate algorithmic narratives into training programs, performance rankings, aesthetic and behavioral standards, spatial arrangements, and financial dependencies. Through these mechanisms, algorithmic uncertainty is rendered manageable, and responsibility for success or failure is transmitted onto individual streamers through disciplined labor practices.}

\newadd{In response to \textbf{\textit{RQ3}}, we show that live-streamers’ reactions to MCNs’ algorithmic interpretations unfold along a patterned trajectory. While many initially comply with MCN guidance as a pragmatic response to uncertainty; others develop doubt as promised outcomes fail to materialize; and some ultimately disengage from MCNs or exit the live-streaming industry altogether. These reactions illustrate how algorithmic interpretations are variably accepted, questioned, and resisted over time.}

\newadd{Thus, this study makes three contributions to CSCW and HCI research:}

\begin{itemize}
\item \newadd{First, it foregrounds intermediary organizations as key sites where organizational algorithmic authority is constructed, influencing how algorithmic knowledge is constituted and operationalized in platform economy.}
\item \newadd{Second, it demonstrates how discursively constructed algorithmic interpretations operate as instruments of organizational labor management. }
\item \newadd{Third, it offers design and policy implications by highlighting the role of intermediary organizations in mediating algorithmic authority and governing algorithmic labor.}
\end{itemize}

\section{Background and Related Work}
\newadd{First, we examine related research on algorithm interpretations, such as folk theories, mental models, algorithmic imaginary, and discursive construction of algorithms (\autoref{folk}).} Second, we analyze the shift from direct algorithmic management (e.g., task allocation or performance metrics) to the softer and more affective forms of control (\autoref{mana}).  Third, we focus on reviewing literature on Multi-Channel Networks (MCNs) and surrounding studies on the live-streaming industry (\autoref{MCN}). 

\subsection{\newadd{Interpretive Frameworks of Algorithms}} \label{folk}
\newadd{This section reviews research that examines how algorithms are understood in everyday practice. Across HCI, CSCW, and digital studies, scholars have developed a range of conceptual frameworks to capture these interpretive processes, including folk theories of algorithms, mental models, algorithmic imaginaries, and recent work on the discursive construction of algorithms. These frameworks foreground how algorithmic systems acquire meaning, shaping how users orient their actions under conditions of algorithmic opacity and uncertainty.}

\subsubsection{\newadd{User-Centered Interpretations of Algorithms}} \label{in}

\newadd{A substantial body of research has examined how workers develop practical understandings of algorithmic systems through experience. Within this literature, several closely related concepts have been used to capture different dimensions of user-centered interpretation. Folk theories of algorithms describe the informal, experience-based explanations that users develop to account for algorithmic behavior, often drawing on trial and error, anecdotal evidence, and peer exchange \cite{devito2017algorithms, dogruel2021folk, huang2022folk}. Mental models emphasize the internal representations users form about how algorithmic systems function, highlighting how these cognitive structures guide expectations and interaction with complex systems \cite{ngo2020exploring}. Algorithmic imaginaries foreground the affective, cultural, and symbolic dimensions of algorithmic understanding, attending to how algorithms are imagined within broader social meanings \cite{bucher2018if}.}

\newadd{Empirical studies show that such understandings often circulate within peer networks, where workers exchange interpretations, compare outcomes, and refine shared explanations of algorithmic behavior \cite{mcdaid2023algorithmic, yao2021together, christin2020metrics, poon2021computer}. Through these processes, interpretations of algorithms become stabilized and inform everyday decisions about what to produce, when to work, and how to respond to algorithmic feedback \cite{bonini2024algorithms, lei2021delivering}. Research on platform labor further demonstrates how these interpretive resources guide routine work practices across domains such as content creation, moderation, and ride-hailing, even when the underlying explanations remain speculative \cite{bishop2019managing, bishop2020algorithmic, cameron2021this, devito2022transfeminine}.}

\newadd{Recent studies also highlight the utility of these interpretive frameworks. For example, Mayworm et al. \cite{mayworm2024content} show how users mobilize shared explanations of content moderation algorithms as diagnostic tools, using them to adjust strategies, anticipate enforcement, and decide when disengagement becomes necessary. Across these accounts, user-centered interpretations of algorithms function as practical resources that render opaque systems actionable in everyday work.}

\subsubsection{\newadd{Discursive Construction of Algorithms and Institutional Mediation}} \label{or}
\newadd{Alongside user-centered and labor-focused approaches, an expanding body of scholarship examines how algorithmic understandings are shaped through institutional discourse and communicative practices. Situated within the broader tradition of algorithmic imaginaries, these studies attend to the ways algorithms are publicly articulated and explained through policy statements, business transparency reports, and educational materials produced by platforms, designers, and regulatory institutions \cite{ananny2018seeing, cellard2022algorithms, schulz2023new}. Rather than focusing on individual sensemaking alone, this line of work traces how authoritative accounts of algorithms are assembled and circulated through formal channels, contributing to shared expectations about how algorithms function and what forms of behavior they reward or penalize.}

\newadd{A key theoretical contribution to this literature is offered by \citeauthor{lin2025vernacular} (\citeyear{lin2025vernacular}), who introduces the concept of the “discursive construction of algorithms” to systematize these institutional processes. Lin \cite{lin2025vernacular} conceptualizes algorithms as objects whose meanings are actively produced through discourse. As institutional actors increasingly publish accounts of algorithmic operations, algorithms become legible through standardized vocabularies that foreground certain principles, causal logics, and evaluative criteria. In this formulation, algorithms are understood as “dynamic, contextual constructs shaped through discourse \cite{lin2025vernacular},” reflecting the situated and contingent ways their meanings are stabilized through communication.}

\newadd{This perspective directs attention to how discursive practices establish interpretive boundaries around algorithms. Lin \cite{lin2025vernacular} summarized that, through selective emphasis, simplification, and abstraction, institutional accounts intentionally highlight particular dimensions of algorithmic systems while leaving others opaque or indeterminate. These processes contribute to the normalization of specific interpretations, shaping what is treated as common sense knowledge about algorithms and which forms of uncertainty are rendered acceptable for users.}

\newadd{Emerging research extends this analysis beyond platform companies to examine how discursive constructions of algorithms travel through intermediary actors that connect institutional narratives with everyday labor practices. Studies of MCNs, fan communities, and other third-party organizations illustrate how algorithmic explanations are translated into optimization advice and even collective action frameworks that orient user behavior \cite{siciliano2023intermediaries, xiao2024let}. At the same time, existing studies provide limited empirical insight into how intermediary organizations integrate discursively constructed algorithmic accounts into labor management.}

\subsection{Algorithmic Management} \label{mana}
Algorithmic management refers to the use of computational systems to allocate, supervise, and evaluate labor, increasingly replacing or supplementing traditional managerial functions in digital and platform-mediated work environments \cite{kellogg2020algorithms,mcdaid2023algorithmic,mohlmann2021algorithmic}. Early scholarship in this field has focused on how platforms use algorithms to automate essential managerial procedures, such as task assignment, performance evaluation, and behavior discipline, especially in gig economy sectors like ride-hailing, delivery, and content moderation \cite{meijerink2023duality,momin2023lumpen}. These studies highlight how algorithmic opacity and asymmetrical access to data reinforce structural inequalities, subjecting workers to rigid systems with limited transparency \cite{rosenblat2016algorithmic,christin2020metrics}. Algorithms, in this view, become infrastructure for managerial control, embedding hierarchical logics in seemingly neutral technical systems.

Recent work has expanded the analytical lens beyond technical automation to include the social, affective, and discursive dynamics of algorithmic management. Scholars have documented how platforms govern not only through direct spoken commands but soft control, i.e., the subtle shaping of worker behavior via gamified dashboards, reward systems, performance metrics, and motivational narratives \cite{cameron2020rise,jarrett2022digital,wood2023platforms}. These systems encourage workers to internalize platform values such as meritocracy, efficiency, and visibility, often prompting aspirational self-discipline under conditions of precarity. This turn in the literature emphasizes how algorithmic power operates at the level of subject formation, producing compliant, hopeful, and responsibilized workers even in the absence of overt coercion \cite{dasgupta2025social,cameron2021seems}.

However, much of this research has concentrated on the direct relationship between platforms and workers, with comparatively less attention to how intermediary institutions, such as third-party service providers, outsourcing firms, or managerial brokers mediate algorithmic systems. These actors have been increasingly serving as epistemic and organizational bridges between opaque platforms and fragmented labor forces, playing a significant yet under-theorized role in shaping how algorithmic control is enacted on the ground.

\subsection{MCNs and Live-Streamers} \label{MCN}

This study investigates how interpretations of algorithms are strategically constructed by intermediary organizations within China’s live-streaming industry. We focus on live-streaming industry as a particularly consequential site: streamers operate in highly volatile algorithmic environments, face extreme uncertainty in monetization, and often lack the technical knowledge to interpret the systems that shape their visibility and income \cite{sun2023unraveling, yao2023crowding, hou2020factors, tiktok2022gifts}. These epistemic conditions make streamers especially dependent on externally provided explanations of how platform algorithms “work”, creating context for the emergence and manipulation of algorithmic interpretations.

Unlike traditional gig work structured around discrete tasks, live-streamers have to continuously perform for audiences in real time, often late into the night, while adapting their appearance, affect, and persona to meet platform and audience expectations \cite{freeman2020streaming, momin2023lumpen, ye2023navigating}. The platform environment further exacerbates this instability. Streamers have to rely on opaque algorithmic recommendation systems to gain visibility, yet they have little security, knowledge or control of how these systems function \cite{sun2023unraveling, yao2023crowding}. Success is contingent on being surfaced by algorithms, but algorithmic volatility means that visibility is uneven and unpredictable. Income mechanisms are primarily gift-based: most streamers do not receive a base salary, and instead depend on voluntary tips or fan gifts from the audience, meaning the digital transactions function more like micro-donations than stable compensation \cite{hou2020factors, sun2023unraveling}. With platforms retaining a significant portion of these gifts, and no obligation for viewers to contribute, streamers are left to bear the risks and uncertainties of monetization \cite{tiktok2022gifts}. Many live-streamers, particularly in the Global South or lower-income regions, enter this labor market with limited digital literacy and educational resources \cite{momin2023lumpen, xiao2024happens, tang2022dare}. They often lack the technical background to interpret algorithmic behavior, analyze platform metrics, or optimize engagement strategies. This limited algorithmic understanding further amplifies their precarity, making streamers vulnerable to misinformation, overwork, self-blame, and self-discipline when platform performance falters \cite{uttarapong2021harassment, ye2023navigating}. 

In this context, MCNs have become central to the construction of algorithmic knowledge and the mediation of labor, which has been a great case for our study in the discursive construction of algorithms. Originating in Western markets such as YouTube, the MCN model initially offered monetization support, branding strategy, and analytics services to individual creators and channel collectives \cite{lobato2016cultural, postigo2016socio}. While the influence of MCNs on YouTube has fluctuated, research has shown that these entities helped consolidate and industrialize early creator labor by aggregating talent, mediating intellectual property claims, and standardizing aesthetic and technical norms across channels \cite{cunningham2019creator,chen2024exploring}. Even when not formally embedded within platform governance, MCNs in the West have played significant roles in disciplining creative labor through soft control mechanisms such as branding scripts, content quotas, and algorithm-focused consultations \cite{siciliano2023intermediaries}.

We selected China as our main context to focus based on it's rapidly developing live-streaming industries and burgeoning intermediary companies. In China, MCNs have become more structurally embedded within the operational logic of platforms such as Douyin, Kuaishou, and Huya. In the meantime, MCNs in China have been institutionalized into platform governance frameworks, actively assisting platforms in managing, monitoring, and optimizing the production of live-streamers \cite{zhang2024contesting, liang2024manufacturing, liu2023zhibo,wang2024harnessing}. Unlike YouTube and other western live-streaming platforms, where most creators remain relatively autonomous ~\cite{alexander2018youtube}, Chinese live-streaming platforms rely heavily on MCNs to coordinate recruitment, training, and day-to-day performance evaluation. As a result, independent live-streamers experience higher competition, and platform labor mediated through MCNs functions both as infrastructural supports and informal managers \cite{gardner2016whats}. These MCNs often promote their technical and analytic capacity to “manufacture” success by decoding and exploiting algorithmic mechanisms, branding themselves as algorithm-savvy “talent incubators” \cite{zhang2024contesting, zhang2024intermediated}.

MCNs act as interpretive infrastructures in the algorithmic era. Their authority stems not from employment arrangements but from their ability to decode opaque platform systems, train streamers to meet algorithmic expectations, and define what counts as desirable labor. Zhang and Tong recently offered a systematic account of the historical development of the MCN ecosystem in China and argued that Chinese content platforms highly rely on MCNs to absorb the risks of professional content production ~\cite{zhang2024contesting}. They also acknowledge that MCNs play an active role in helping creators become familiar with platform algorithms and optimize their content accordingly ~\cite{zhang2024contesting}. However, there is still a gap in literature on investigating critical questions: how do MCNs specifically construct and disseminate knowledge about algorithms, and why do they do so in particular ways? Thus, our research aims to explore the role of MCNs in shaping creators’ epistemic relationship to platform algorithms as both interpreters and enforcers.

\section{Methods}
In this study, we conducted extensive fieldwork primarily in Beijing and Changsha in China, focusing on MCN companies and live-streamers. From April to December, 2023, we engaged in a nine-month in-depth ethnographic investigation. Our fieldwork involved observing the daily operations of three MCN companies (\newadd{A, B, C in our cases})—pseudonyms used to protect the anonymity of participating organizations—including how they set up live-streaming environments, provided technical and emotional support, and managed interactions among managers, technologists, and live-streamers. We also followed streamers’ broader daily routines to understand their labor conditions and platform practices in context. All observations were conducted with the consent and cooperation of both the MCN companies and the individual participants.

To supplement this ethnography, we conducted one-on-one online interviews with 24 live-streamers, \newadd{seven former live-streamers,} and 13 MCN practitioners. While our immersive fieldwork was based in three MCNs, our interview sample expanded to include participants from a total of seven MCN companies \newadd{(A, B, C, D, E, F, G)}.

This research received approval from the university’s Institutional Review Board (IRB). Throughout the study, we were transparent about our researcher roles and obtained informed consent from all participants. 

\subsection{Participants: Live-Streamers (N=24), \newadd{Former Live-Streamers (N=7)}, MCNs Practitioners (N=13)}

We interviewed 24 live-streamers (L1–L24) active on major Chinese platforms including Laifeng, MoMo, YY, Douyin, and Huajiao. Participants ranged in age from 21 to 34 and came from diverse gender, educational, and geographic backgrounds (see \autoref{tab:my_label1}). About two-thirds (n = 15) were full-time streamers, while the remainder worked part-time, supplementing their unstable streaming income with jobs in nail salons, barbershops, massage parlors, supermarkets, electronics factories, or as fitness instructors, models, and salespeople. We also interviewed 13 MCN practitioners (M1–M13) from seven companies spanning different stages of organizational development, including startups, growth-stage firms, and mature MCNs (see \autoref{tab:my_label2}). Their roles included executives (n=5), managers (n=6), and technicists (n=3), with one participant holding a hybrid manager-technician position. Executives made strategic decisions; managers coordinated scheduling and performance; and technicists provided technical support and training. \newadd{We also interviewed seven former live-streamers (F1–F7) who had previously been affiliated with MCNs and had left the live-streaming industry (see \autoref{tab:my_label3}). } \newadd{It is noted that F1 and F7 initially left their MCNs to stream independently without any MCNs for a while before ultimately exiting the industry. F5 had switched between two MCN companies prior to departure.} 

\newadd{As of December 2023, approximately 15.08 million individuals in China have adopted live-streaming as their primary occupation, with 57.1\% citing economic income as their main motivation for entering the industry. 68\% of professional live-streamers report average monthly incomes below 3,000 RMB (Approximately \$420) \cite{people2024livestreamer, cpla2024livestreamer}.}

\begin{table}[h!]
    \centering
    \caption{Information of Live-Streamer Participants. }
    \label{tab:my_label1}
    \footnotesize
    \setlength{\tabcolsep}{3.5pt}
    \begin{tabular}{lcccccccccc}
        \toprule
        Name & Gender & Age & MCN & Work Tenure & Education & Platform & Fans & Employment & Income & Domain\\
        \midrule
        L1 & Woman & 21 & A & 8 months & Undergraduate & Laifeng & 3k & Part-time & ¥1000-2000 & Chat \\
        L2 & Man & 24 & A & 6 months & Master & Laifeng & 2k & Part-time & ¥1000-1500 & Chat \\
        L3 & Woman & 21 & D & 3 months & Undergraduate & Laifeng & 1k & Part-time & ¥3000-4000 & Music \\
        L4 & Woman & 25 & B & 3 months & Master & Laifeng & 1k & Part-time & ¥2000-3000 & Music \\
        L5 & Woman & 22 & C & 7 months & Senior high school & Laifeng & 23k & Full-time & ¥2000-4000 & Music \\
        L6 & Woman & 26 & C & 8 months & Senior high school & Laifeng & 1k & Full-time & ¥1500-2500 & Dance\\
        L7 & Man & 24 & B & 3 months & Junior college & Laifeng & 1k & Full-time & ¥1500-3000 & Chat \\
        L8 & Woman & 24 & B & 9 months & Junior college & Laifeng & 1k & Part-time  & ¥2000 & Chat \\
        L9 & Woman & 18 & D & 2 years & Junior college & Laifeng & 6k & Full-time  & ¥3500-4000 & Chat \\
        L10 & Woman & 23 & C & 5 years & Senior high school & Laifeng & 148k & Full-time & ¥10000-15000 & Music\\
        L11 & Woman & 26 & C & 4 years & Undergraduate & Laifeng & 98k & Full-time & ¥10000-12000 & Music\\
        L12 & Woman & 28 & C & 2 years & Senior high school & Laifeng & 57k & Full-time & ¥8000-10000 & Dance \\
        L13 & Woman & 20 & G & 3 years & Junior college & MoMo & 12k & Full-time & ¥3000-6000 & Chat \\
        L14 & Woman & 28 & G & 1 year & Senior high school & MoMo & 1k & Full-time & ¥4000-4500 & Chat \\
        L15 & Woman & 26 & G & 1.5 years & Junior college & MoMo & 1k & Full-time & ¥3000-5000 & Chat \\
        L16 & Woman & 24 & F & 2.5 years & Junior college & YY & 2k & Full-time & ¥1000-3500 & Music \\
        L17 & Woman & 28 & F & 1 year & Undergraduate & YY & 1k & Full-time & ¥2000-4000 & Music \\
        L18 & Woman & 23 & C & 1 year & Undergraduate & Douyin & 2k & Full-time & ¥1500-2000 & Music \\
        L19 & Man & 21 & C & 1.5 years & Undergraduate & Douyin & 83k & Part-time & ¥2000-3000 & Music \\
        L20 & Man & 24 & F & 2 years & Undergraduate & Douyin & 75k &  Part-time & ¥3000-4000 & Fitness \\
        L21 & Man & 34 & E & 3 months & Undergraduate & Douyin & 2k & Part-time & ¥1000-1500 & Music \\
        L22 & Man & 26 & E & 4 months & Master & Douyin & 9k & Part-time & ¥2000-2500 & Chat \\
        L23 & Man & 29 & F & 2 years & Junior college & Douyin & 15k & Full-time & ¥4000-5000 & Dance \\
        L24 & Woman & 32 & F & 1 year & Master & Huajiao & 3k & Full-time & ¥3000-3500 & Music \\
        \bottomrule
    \end{tabular}
\end{table}

\begin{table}[h!]
    \centering
    \caption{Information of MCNs Participants}
    \label{tab:my_label2}
    \footnotesize
    \begin{tabular}{lcccccccc}
        \toprule
        Name(ID) & MCNs & MCNs Stage & Position & Gender & Age & Working Seniority & Education \\
        \midrule
        M1 & B & Mature & Executive & Man & 42 & 4 years & Master \\
        M2 & D & Startup & Manager & Woman & 35 & 2 years & Junior college \\
        M3 & A & Mature & Executive & Woman & 42 & 5 years & Undergraduate \\
        M4 & C & Growth & Manager/Technicist & Man & 40 & 4 years & Junior college \\
        M5 & A & Mature & Manager & Woman & 40 & 1 year & Undergraduate \\
        M6 & C & Growth & Executive & Man & 48 & 5 years & Junior college \\
        M7 & F & Startup & Executive & Woman & 46 & 2 years & Ph.D \\
        M8 & E & Startup & Manager & Man & 53 & 2 years & Undergraduate \\
        M9 & G & Mature & Manager & Man & 25 & 2 years & Junior college \\
        M10 & F & Startup & Manager & Woman & 35 & 2 years & Master \\
        M11 & G & Mature & Executive & Man & 30 & 3 years & Junior college \\
        M12 & C & Growth & Technicist & Man & 24 & 1 year & Junior college \\
        M13 & C & Growth & Technicist & Man & 22 & 6 months & Undergraduate \\
        \bottomrule
    \end{tabular}
\end{table}

\begin{table}[h!]
    \centering
    \caption{\newadd{Information of Former Live-streamers Participants. }}
    \label{tab:my_label3}
    \footnotesize
\setlength{\tabcolsep}{3.5pt}
    \begin{tabular}{lcccccccccc}
        \toprule
        Name & Gender & Age & MCN & Work Tenure & Education & Platform & Fans & Employment & Income & Domain\\
        \midrule
        F1 & Woman & 22 & / & 1 year & Undergraduate & Douyin & 34k & Full-time & ¥4000-5000 & Chat \\
        F2 & Woman & 31 & A & 6 months & Undergraduate & Laifeng & 2k & Part-time & ¥2000-3000 & Chat \\
        F3 & Woman & 28 & F & 1.5 years & Senior high school & YY & 6k & Full-time & ¥3000-5000 & Music \\
        F4 & Woman & 26 & F & 8 months & Senior high school & YY & 5k & Full-time & ¥3000-3500 & Dance \\
        F5 & Man & 29 & G & 2 years & Senior high school & Momo & 23k & Full-time & ¥4000-5000 & Chat \\
        F6 & Man & 23 & E & 6 months & Junior college & Douyin & 52k & Part-time & ¥1000-3000 & Fitness \\
        F7 & Man & 21 & / & 6 months & Undergraduate & Douyin & 28k & Part-time & ¥2000-3000 & Fitness \\
        \bottomrule
    \end{tabular}
    \vspace{0.5em}
    \raggedright
    \footnotesize
    \textit{Note.} Live-streaming data reflects account information at the time of departure from the industry.
\end{table}

\subsection{Study Design}

\subsubsection{Nine-month Ethnography in Beijing and Changsha}

We conducted a nine-month ethnographic study (April–December 2023) within three MCNs \newadd{(A, B, C)} based in Beijing and Changsha, actively participating in their live-streaming practices. These two cities were chosen because they represent distinct yet influential hubs in China’s live-streaming economy: Beijing, as the nation’s capital, is a center of digital entrepreneurship and policy discourse \cite{cunningham2019china}; Changsha, though a smaller city by comparison, has emerged as a leading cultural production hub that attracts a large number of rural migrant streamers due to its thriving live-streaming industry \cite{xiao2024happens}. 

Over the course of our nine-month ethnographic fieldwork, we focused on how MCNs and live-streamers interact around platform algorithms. In the first three months, we took an open and exploratory approach to learn how streamers get started, what their daily routines look like, and how MCNs provide support or step in to guide their work. During this period, we also picked up basic streaming skills ourselves and built trust with both streamers and MCN staff. In the middle phase of our fieldwork, we started noticing how often both MCNs and streamers talked about “the algorithm.” The algorithm became a key part of everyday practice. MCNs used it to set performance expectations, give advice, and explain success or failure. Streamers, in turn, developed their own informal beliefs about how the algorithm works, often based on MCN coaching. In the final four months, we shifted our attention fully to this process. We closely examined how MCNs interpret vague and unpredictable algorithmic signals and turn them into concrete rules and routines for streamers to follow. This shift led us to understand MCNs both as business managers and as algorithmic intermediaries that translate platform uncertainty into structured labor expectations.

\subsubsection{Interviews with Live-Streamers and MCNs Practitioners}
Following our nine-month ethnography, we conducted semi-structured interviews in the end of our ethnography with 24 live-streamers and 13 MCN practitioners to further investigate how MCNs mediate algorithmic dynamics.

Participant recruitment began with the three MCNs involved in our fieldwork, who helped us access a broader network of streamers and affiliated MCNs, to our final sampled seven MCNs, allowing us to diversify our sample beyond the initial field sites. This expansion was necessary for two reasons. First, MCNs vary considerably in size, specialization, and degrees of platform dependency, which can lead to various algorithmic strategies and labor practices. Second, broadening the sample helped us identify cross-organizational common patterns, enabling a more robust understanding of how algorithmic knowledge is constructed and circulated across the MCN ecosystem. We intentionally included both novice and experienced streamers, as well as MCN staff from different roles and organizational levels.

The interview protocol, developed based on prior research on live-streamers~\cite{wohn2020audience, wang2019love, wu2023streamers,chen2024exploring} and MCNs~\cite{vonderau2016video,liang2022end,siciliano2023intermediaries}, then focused on several core areas: challenges associated with platform algorithms, streamers’ understanding and navigation of algorithmic systems, the forms of algorithmic guidance or control exerted by MCNs, and the infrastructural or interpretive support MCNs provide to facilitate compliance or success. For live-streamers, sample questions included: \textit{“How do you know when the algorithm is working in your favor or against you?”}; \textit{“Has your MCN ever provided you with explanations or advice about the algorithm?”; and “What kind of support—technical, emotional, or interpretive—do you receive from your MCN in navigating platform challenges?”} For MCN practitioners, we asked: \textit{“How do you stay informed about changes in the platform’s algorithmic rules?”}; \textit{“Do you provide training or informal guidance to streamers on how to ‘play to the algorithm’?”}; and \textit{“Would you describe your role for live-streamers regarding platform algorithms?”}

Each interview lasted approximately 60 to 90 minutes. Throughout, we encouraged participants to critically reflect on how platform algorithms shape their everyday labor and how MCNs influence these experiences as organizational actors and mediators of algorithmic knowledge and expectations. 

\subsubsection{\newadd{Interviews with Former Live-Streamers}}

\newadd{To address potential survivor bias in studies of platform labor, after our ethnography and interviews with streamers and MCNs staffs, we also conducted semi-structured interviews with seven former live-streamers (F1–F7) who had previously been affiliated with MCNs and had exited the live-streaming industry (see \autoref{tab:my_label3}). We realized accounts from active and relatively successful workers may overrepresent perspectives shaped by continued participation and organizational alignment, while obscuring experiences of disillusionment. Including former live-streamers enabled us to examine how MCN practices and algorithmic interpretations were reassessed after participants were no longer embedded in ongoing managerial relationships. Each interview lasted approximately 60 to 90 minutes.}

\newadd{Former live-streamers provided retrospective accounts that spanned both their period of active participation and their post-exit reflections. These interviews focused on how MCNs explained platform algorithms, how such explanations shaped streamers’ expectations and labor decisions, and how participants evaluated these narratives after leaving the industry. Sample questions included: \textit{“Looking back, how do you now interpret the advice and explanations provided by your MCN?”}; and \textit{“What role did algorithm-related expectations and MCNs you met play in your decision to leave?”} }

\subsubsection{\newadd{Reflection of Study Design}}
\newadd{Given the evaluative and hierarchical nature of MCN-managed live-streaming work, we remained attentive to how organizational conditions might shape participants’ accounts during ethnographic observation and interviews. We noticed that MCNs exercise varying degrees of control over streamers’ income, visibility, and access to future opportunities, which can influence how experiences are narrated in research settings. We also recognized that some MCN employees might be reluctant or hesitant to articulate critical reflections on organizational practices, particularly when such discussions could be perceived as professionally sensitive.}

\newadd{To ensure the reliability of our findings, we adopted several complementary methodological strategies. First, we emphasized confidentiality as well as the separation of research activities from organizational evaluation or performance assessment processes during interviews. We encouraged participants to begin with detailed descriptions of everyday work routines, which allowed accounts of algorithmic guidance and MCNs practices to emerge through situated examples.}

\newadd{Second, we conducted a relatively extensive set of interviews, which allowed recurring themes to emerge across accounts. Our interviewees included MCN staff in relatively senior positions like executive, who were less constrained by immediate power dependencies and more willing to speak candidly about organizational strategies. We also interviewed former streamers who no longer maintained contractual or financial ties with MCNs, reducing concerns about retaliation or reputational risk. These interviews provided perspectives that were less shaped by ongoing dependency relationships.}

\newadd{Third, we relied on systematic triangulation across multiple data sources. Interview data were examined in conjunction with long-term ethnographic observations, informal conversations, internal training materials, and accounts from MCN practitioners, streamers, and former streamers.} \newadd{The longitudinal nature of our ethnography also provided repeated opportunities to observe how algorithmic narratives were enacted in practice, reinforced through training and evaluation, and reproduced across everyday interactions.}

\subsection{Data Analysis}
All interviews were conducted and recorded in Chinese to ensure participants could communicate comfortably and effectively. Three researchers subsequently transcribed and translated these recordings into English for analysis. We conducted a thematic analysis, combining both deductive and inductive coding methods. Deductive codes were informed by our theoretical framework and interview protocol, centering on constructs such as algorithmic governance, mediation, risk transfer, and infrastructural support. Inductive codes emerged through open reading of the transcripts and fieldnotes, capturing unanticipated practices, metaphors, tensions, and expressions from participants. We analyzed our ethnographic fieldnotes and interview transcripts together, rather than treating them as separate datasets. This integrated approach allowed us to triangulate across different forms of data, what participants said in interviews and what we observed in situ, yielding richer contextualization and interpretive depth ~\cite{xiao2024let}. Drawing from traditions in interpretive ethnography and qualitative HCI, we treated interviews as situated reflections shaped by ongoing organizational and algorithmic conditions observed during fieldwork ~\cite{seaver2017algorithms}. 

We adopted a reflexive thematic analysis \cite{braun2019reflecting}, following the traditional qualitative HCI approach ~\cite{kuo2023understanding,xiao2024let,xiao2025might}, which focuses on the interpretive work of the researchers and views knowledge production as iterative and situated. To begin, all three researchers independently coded a subset of transcripts and fieldnotes to generate initial codes. These were compared in regular team discussions, where we negotiated code definitions, added emergent categories, and refined our shared codebook. Rather than calculate inter-coder reliability, we emphasized interpretive alignment through dialogue, clarification, and documentation of divergent readings, following qualitative HCI best practices. After completing the open coding phase, we clustered related codes into higher-order conceptual categories. We used affinity diagramming~\cite{haskins2020using} for organizing codes, identifying cross-cutting themes, and mapping connections between concepts. 

\subsection{\newadd{Positionality Statement}}

\newadd{Our research team brings situated perspectives shaped by years of engagement with China’s live-streaming ecosystem through research and industry experience. Our team members conducted extensive research on platform labor and digital economies, including several  published paper on MCNs, live-streamers and platform policy. This background informed our sensitivity to everyday labor routines and the informal norms through which platform work is organized.}

\newadd{The research team’s engagement with MCN companies enabled access to organizational meetings, training sessions, and internal documents that would otherwise be difficult to observe. During the course of this engagement, one member of the research team contributed professional technical reports and analytical materials for multiple MCN companies, work that led to sustained collaboration and, at one point, an invitation to join an MCN as a formal employee. This experience provided firsthand exposure to internal processes, performance evaluation logics, and the organizational circulation of algorithm-related knowledge. Such role immersion are also characteristic of ethnography, in which researchers gain access and insight through sustained involvement in organizational practice \cite{dumont2023immersion}.}

\newadd{At the same time, such proximity required ongoing reflexive attention to power asymmetries and role boundaries, particularly when interpreting managerial narratives and performance-oriented accounts. Throughout the study, organizational perspectives were examined alongside ethnographic observations and interviews with active and former streamers. This analytic triangulation helped ensure that institutional viewpoints were situated within a broader field of perspectives and did not dominate the interpretation of algorithmic practices and labor relations.} \newadd{We engaged in regular reflexive discussions to examine how prior field relationships and assumptions about platform work shaped our research. These discussions informed the refinement of interview protocols and the articulation of key analytical concepts.}

\section{\newadd{Findings: MCNs’ Discursive Construction of Algorithms (RQ1)}}\label{RQ1}

Live-streaming platforms rarely disclose how their recommendation systems function. As a result, live-streamers must navigate a system that is unpredictable, constantly shifting, and difficult to decipher. For streamers, this opacity often leads to frustration and powerlessness. They cannot understand why one stream gains traction while another goes unnoticed, or why their platform visibility drops without warning. \newadd{This uncertainty is particularly acute for streamers with limited formal education or prior work experience, a common profile within the live-streaming workforce. Lacking alternative sources of technical or professional guidance, these streamers often rely heavily on MCNs’ explanations of how algorithms work and how exposure can be improved. MCNs thus become a primary interpretive authority through which algorithmic outcomes are understood.}

\newadd{At the same time, during our ethnography and interviews, we found that MCNs themselves do not occupy a privileged insider position. Like the streamers they manage, MCNs also lack access to the internal mechanism of platform algorithms and must operate through inference, observation, and accumulated experience.} To make sense of this algorithmic uncertainty, MCNs observe performance trends, compare outcomes across multiple streamers, and develop related algorithmic “knowledge”: informal, experience-based beliefs about how the algorithm behaves, what kinds of content it rewards, and how it might be influenced. They are grounded in trial-and-error and pattern recognition, not in verified mechanisms. \newadd{Guidance derived from these beliefs is widely understood within MCNs as having the potential to improve algorithmic visibility at an aggregate level. MCN practitioners often described such advice as \textit{“generally helpful” } (M6) or \textit{“likely to work for most streamers”} (M7), while acknowledging its limits in practice. They could not determine whether a given recommendation would be effective for any specific streamer, nor could they specify the magnitude or durability of its effects. As many MCN interviewees in our study admitted, there is no formula that ensures algorithmic success, even when streamers closely follow the advice they are given.} \newadd{Despite this internal shared uncertainty, MCNs often present their algorithmic explanations with confidence and clarity when communicating with streamers and instructing streamers how to deal with algorithms. This asymmetry between internal uncertainty and external certainty forms the basis for how MCNs construct and disseminate algorithmic interpretations within live-streaming industry. We thus focus on how MCNs intend to construct their algorithmic interpretations in this paper.} 

\newadd{Here, we decide to use the term \textit{discursive construction of algorithms} to characterize MCNs’ algorithm-related explanations because these accounts function as intentional and strategic interventions to streamers’ behavior. Unlike folk theories, which typically emerge as users’ folk attempts to understand opaque systems, MCNs’ algorithm explanations are developed and deployed at organizational level. They are crafted to coordinate behavior across multiple streamers, standardize content production practices, and justify managerial directives. Algorithmic explanations become tools for governing labor rather than personal interpretations held by individual users. Similarly, while the concept of algorithmic imaginaries emphasizes how users envision algorithms and their social implications, MCNs’ practices extend beyond imagination. Their discursive constructions selectively amplify certain algorithmic attributes, such as rewardability, while downplaying uncertainty. This selective emphasis serves concrete organizational goals, particularly the management of streamer labor and the pursuit of increasing profitability.} 

The discursive construction of algorithms by MCNs includes internal and external interpretations to manage labor and increase profits. In the sections that follow, we show how MCNs construct these two interpretations—one inward-facing and probabilistic (\autoref{internal}), the other outward-facing and prescriptive (\autoref{external}). \newadd{This \autoref{RQ1} here focuses on the substance of these interpretations, that is, what MCN staff and streamers come to believe about how the algorithm works within this explanatory framework. In following \autoref{RQ2}, we extend this analysis by examining how these algorithmic interpretations are mobilized in labor management practices.}

\subsection{Internal Algorithmic Interpretation: Managing Uncertainty and Risk within MCNs} \label{internal}

\newadd{As mentioned by M1, the executive of a famous MCN, MCN’s algorithmic understandings are initially hoped to be oriented toward increasing the algorithmic visibility of affiliated streamers, thereby generating higher traffic, longer viewing times, and ultimately greater live-streaming revenue. Improved streamer performance translates into higher commission income, stronger bargaining power with platforms, and greater profitability for MCNs themselves. To achieve this goal, MCNs accumulated operational experience and the circulation of informal industry knowledge. Staff continuously test different content formats, streaming schedules, and interaction styles across multiple streamers, while also exchanging observations through personal networks, platform rumors, and peer organizations. These diverse sources are synthesized into actionable understandings of how algorithmic visibility might be advanced under conditions of opacity. }

\newadd{These internal understandings of algorithms are expressed through heuristic rules and behavioral scripts, often referred to within organizations as “formulas” or “track logics.” Typical examples include recommendations on streaming frequency (e.g., “at least three sessions per week”), minimum broadcast duration (e.g., “streams under 40 minutes won’t get recommended by algorithms”), and emotional tone (e.g., “you must be excited in the first five minutes, or the system will judge it as low-interaction”). Beyond behavioral scripts, many MCNs also develop aesthetic and infrastructural prescriptions that they believe enhance algorithmic visibility. Staff frequently emphasize the importance of high-definition visuals, stable audio quality, lighting conditions, and even the physical arrangement of the streaming environment. These elements are framed as signals that the algorithm is assumed to recognize and reward.}

However, in practice, while these algorithmic interpretations offer a sense of control, most MCN staff readily admit their limitations, no matter how much efforts MCNs made to understand algorithms. \textit{“Even if you follow all the best practices, sometimes the platform just doesn’t give you any algorithmic visibility,”} M9 remarked. This tension between the promise of algorithmic optimization and the reality of its unpredictability has prompted many MCNs to shift focus. Rather than attempting to fully decode the algorithm, MCNs increasingly prioritize monetization logic.

\newadd{MCNs continue to hope to increase the visibility of each affiliated streamer and to provide support toward that goal. At the same time, they are fully aware of the algorithm’s fundamental unpredictability. As a result, MCNs adopt a statistical approach to profitability that treats uncertainty as something to be managed rather than eliminated.} By recruiting large cohorts of new streamers and investing minimal resources in each, MCNs reduce the risk associated with any one individual’s failure. The goal is not to ensure the success of every recruit but to ensure that some will be algorithmically \textit{“picked.”} As M9 elaborated: \textit{“This process is highly random. We recruit newcomers who look attractive, provide them with some equipment, and ask them to start live-streaming. But we don’t know who will make it. What we do know is that if we have enough people, a certain percentage will succeed. Then we allocate resources to those who are ‘chosen’ by the algorithm.”} This batch recruitment model allows MCNs to scale labor while minimizing upfront investment and maximizing return on attention. 

But the logic does not stop at internal risk distribution. Just as MCNs construct algorithmic interpretation internally to manage uncertainty, they also construct simplified algorithmic interpretation externally—targeted at streamers. Unlike internal interpretation, which circulate among MCN staff as tentative, probabilistic tools for interpreting platform behavior, external interpretations are simplified and confident theories constructed for streamers. These externally projected interpretations serve two strategic purposes: first, they attract and retain labor by portraying algorithmic success as predictable and achievable, thereby encouraging newcomers to join and existing streamers to continue; second, they stimulate financial investment by persuading streamers to purchase services, equipment, or training packages under the belief that these will enhance their algorithmic performance, allowing MCNs to profit from these sales.

In this study, we do not seek to fully examine how MCNs develop sophisticated understandings of platform algorithms in order to manipulate them—indeed. In our fieldwork, MCNs rarely invested substantial effort in decoding or directly manipulating the algorithm, which they largely viewed as infeasible, unrewarding, and not worth the effort given the high uncertainty and low return. Rather, our focus lies in how MCNs construct and mobilize discursive construction of algorithms—as organizational narratives—to manage algorithmic uncertainty, structure labor practices, and sustain control over streamers within the live-streaming ecosystem. This is a strategy that, in the context of our sampled Chinese MCNs, represents a more direct and prevalent path to profitability than algorithmic manipulation.

What finally emerges is a dual system of epistemic practice about algorithmic interpretations: internally, MCNs recognize the algorithm as volatile and uncontrollable; externally, they reframe it as predictable and improvable. This asymmetry is not incidental—it is the foundation of the MCN business model. By selectively mobilizing algorithmic interpretations in different contexts, MCNs manage precarity, sustain labor supply, and extend their authority within China’s live-streaming industry. Next, in \autoref{external}, we further analyze how these externally directed algorithmic interpretations are strategically constructed by MCNs for streamers.

\subsection{External Interpretation: Simplified Narrative to Manage and Govern Streamers} \label{external}

Unlike the probabilistic and speculative nature of internal algorithmic interpretations held by MCN staff, these outward-facing interpretations of algorithms targeting streamers are packaged as confident, simplified rules that make the algorithm seem legible and controllable. We identify two primary ways in which these algorithmic interpretations are strategically deployed: (1) by translating algorithmic logic into concrete, observable metrics such as appearance and equipment quality (\autoref{inter}); (2) by framing algorithmic success as attainable through discipline, investment, and compliance (\autoref{frame}). \newadd{Here we focus more on the content of these beliefs and we will further analyze why MCNs constructed these interpretations in such ways for labor management in \autoref{RQ2}.}

\subsubsection{Simplifying Algorithmic Logic Through Observable Metrics} \label{inter}
In the face of algorithmic opacity and instability, MCNs firstly offer simplified but confident explanations to live-streamers that translate the platform’s black-box logic into concrete, observable metrics. For example, M2 usually explains to new streamers that physical appearance plays a decisive role in gaining algorithmic exposure by increasing early viewer retention: \textit{“The audience will only give a live-streamer three seconds to show themselves and keep them. When a live-streamer is particularly attractive, they are more likely to gain audience favor. This results in higher viewer retention and influx rates, leading the recommendation algorithm to classify them as popular live-streamers and, consequently, grant them more exposure.”} This logic, while unverified, is widely internalized by streamers.

\begin{figure}[htbp]
    \centering
    \includegraphics[width=0.8\textwidth]{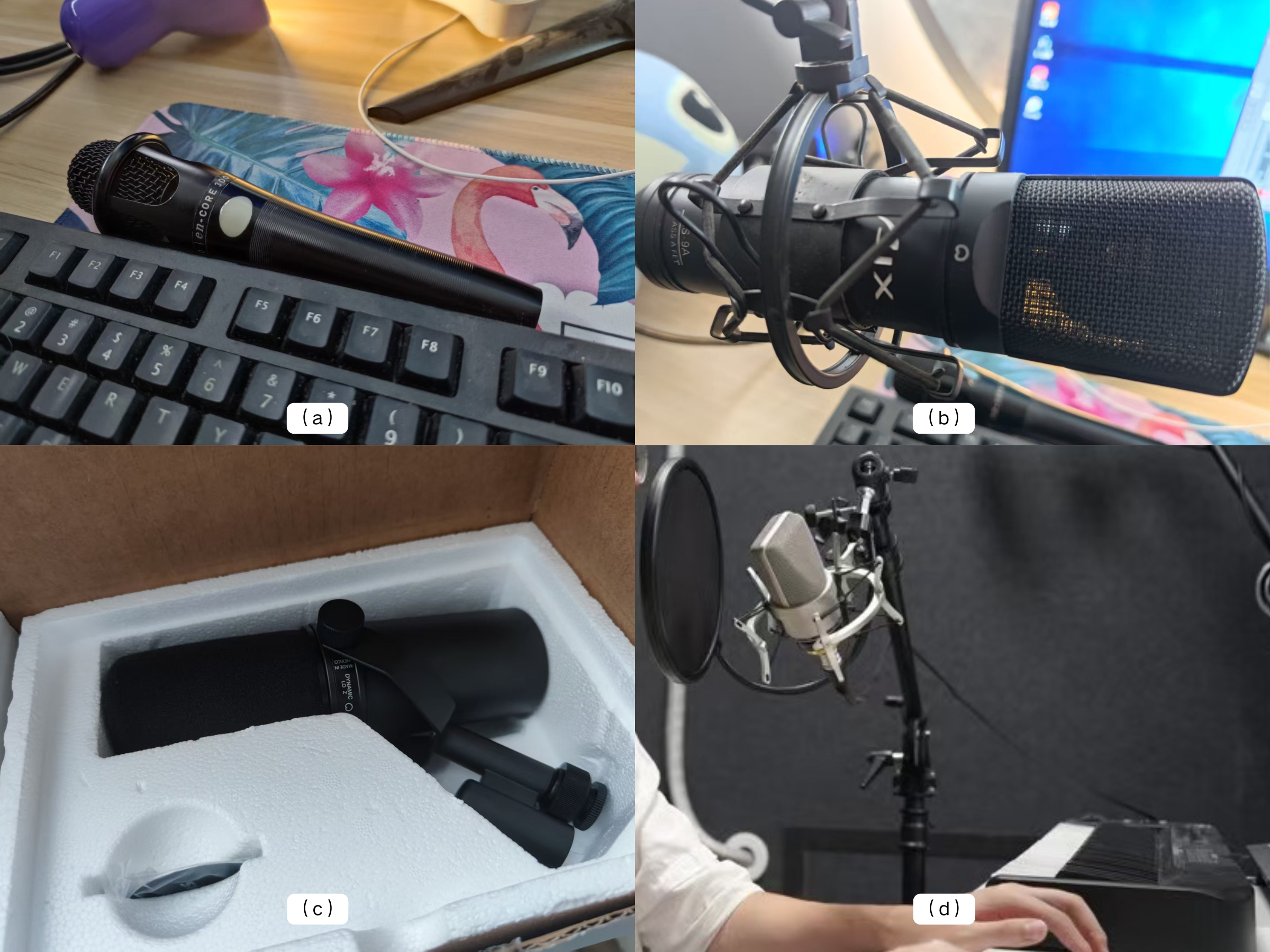}
    \caption{Tiered technical audio solutions deployed by MCN agencies serve as material markers of professional status and algorithmic worth within the live-streaming economy. (a) Entry-level handheld microphone provided as standard equipment for contracted streamers, a counterfeit version of the Blue E300, with production costs ranging from 60-70 RMB (approximately 8 USD). (b) Mid-tier condenser microphone offered complimentary with streaming studios, including free replacement service for malfunctions; this unit replicates the Neumann TLM 193 with manufacturing costs between 300-400 RMB (approximately 50 USD). (c) Shure SM7B microphone recommended by MCN agencies for talk show hosts, voice actors, and vocal performers with established audience bases; streamers are encouraged to purchase this equipment through the MCN at a one-time fee of 4,000 RMB (market value approximately 2,500 RMB), inclusive of complementary mixing console and technical support services. (d) Flagship audio broadcasting system typically reserved for contracted knowledge-based/educational teams or professional singers; streamers may temporarily rent this equipment for cover or original song recordings at approximately 300 RMB (40 USD) per hour, with post-production and technical assistance available at additional cost. This premium equipment primarily serves as a demonstration of the MCN’s professional capabilities and is not available for external purchase.}
    \label{fig:microphone_tier_comparison}
\end{figure}

Another common metric emphasized by MCNs is the quality of equipment. M4 explained that many platforms set technical thresholds for stream quality, and that insufficient equipment could lead to streams being algorithmically filtered out: \textit{“If your sound and visuals are poor, people drop out quickly, and that hurts your retention rate. Algorithms pick up on that. So we always recommend at least the basics, e.g., lights, microphones, stable connection.”} Streamers echoed this perceived necessity. L5 noted that even renting equipment was necessary for basic visibility:\textit{“Whether renting or buying equipment, it at least gives you a chance to gain some recommended followers and earn a bit of money. Without investing in equipment, it’s nearly impossible to be visible to viewers.”} As shown in \autoref{fig:microphone_tier_comparison}, MCNs distribute tiered technical audio equipment in ways that symbolically link production quality with algorithmic visibility, reinforcing an externally imposed algorithmic interpretations that equates financial investment in equipment with success.

What often begins as a recommendation to improve a single technical element, such as audio clarity or lighting, gradually expands into a broader vision of the “ideal” streaming setup. MCNs do not simply promote individual tools; they encourage streamers to view the entire streaming environment as a form of algorithmic currency. Ring lights, DSLR cameras, soundproofed backdrops, multi-angle mounts, branded filters, and LED mood lighting are framed as signs of professionalism and algorithm-readiness. As streamers internalize this logic, the line between technical sufficiency and aesthetic perfection becomes blurred. A basic setup is no longer seen as enough to \textit{“meet the algorithm’s baseline”} (M9); instead, streamers feel pressured to constantly upgrade in order to stay competitive within an ever-rising standard. 

MCNs continue to construct the algorithm as a controllable input-output system: if streamers follow the right steps, if maintain a certain visual style, upgrade equipment, stream for long enough, and use the right tags, they are told they will be rewarded with visibility and growth. This framing simplifies algorithmic complexity into a checklist of behaviors and attributes.

Because this interpretation of algorithm constructed by MCNs feels actionable, most streamers often accept it without question. They internalize the idea that if they fail to gain traction, it is because one of these variables was insufficient. Perhaps their lighting was not professional enough, their background not aesthetic enough, their streaming duration too short, or their performance too flat. This belief system allows MCNs to continuously raise performance benchmarks, always pointing to another metric that could be optimized. 

\subsubsection{Framing Algorithmic Success as Attainable and Controllable} \label{frame}

A second common way MCNs explain the algorithm is by framing it as attainable and controllable, a fair system that rewards effort, discipline, and the right technical setup. This interpretation is actively constructed by MCNs during recruitment and onboarding. New streamers are told that the algorithm can be tamed, and that their success depends on how closely they adhere to MCN advice and technical standards. This framing channels attention away from platform bias or structural inequality, and toward individual action and responsibility.

MCNs often present live-streaming platforms and their algorithmic infrastructures as less discriminatory and more open than other forms of labor or public-facing institutions. In particular, our sampled MCNs in China emphasize that streamers from rural, migrant, or female backgrounds may find more opportunities online than in traditional workplaces. Unlike offline industries where class, gender, accent, or education may limit mobility, MCNs describe algorithmic systems as blind to such characteristics, privileging performance, persistence, and technique instead. Several streamer interviewees (n=7) recalled being told during on-boarding sessions that \textit{“the algorithm doesn’t care who you are; everyone’s talents deserve to be seen”}, a message that resonated deeply with streamers from marginalized backgrounds.

This perception helps explain why live-streaming, despite its precarity, continues to attract disproportionately high numbers of rural migrants and young adults, especially women. The promise of algorithmic neutrality, reinforced by MCNs’ messaging, offers both a psychological appeal and a perceived escape from entrenched discrimination elsewhere.

It is precisely this belief that the algorithm can be tamed sustains repeated investment from streamers. Streamers are told that success is not arbitrary, but a matter of effort. The algorithm, unlike human employers or traditional gatekeepers, is portrayed as objective and responsive by MCNs. For many streamers, especially those excluded from stable or upwardly mobile labor markets, the promise of algorithmic fairness feels both empowering and worth pursuing. This perception justifies the continued purchase of equipment, the adoption of performance routines, and the sacrifice of time and emotional energy. Even when results fall short, the interpretations remains intact: if one stream fails, the next might succeed. 

\section{\newadd{Findings: Labor Management as a Result of MCNs’ Discursive Construction of Algorithms (RQ2)}}\label{RQ2}

Building on \autoref{RQ1}, which examined how MCNs construct internal and external interpretation of algorithms, we now turn to how these interpretations are institutionalized into labor control. First, MCNs embed algorithmic expectations into everyday routines, through on-boarding, coaching, and evaluations, repeating the message that effort, affect, and consistency will be rewarded by the algorithm. This moralized framing aligns with the external interpretation that success is earned, motivating streamers to self-discipline and persist. At the same time, MCNs internally acknowledge the algorithm’s volatility and shift the burden of failure onto streamers through this moral frame, who blame themselves rather than systemic opacity (\autoref{trans}). Second, MCNs formalize these beliefs through structured schedules and aesthetic standards, reinforcing the idea that algorithmic success can be optimized. Streamers are encouraged to upgrade equipment, refine routines, and purchase services—all justified by the external interpretation. Yet MCNs, guided by internal interpretation, understand the algorithm offers no guarantees. Their profits rely not on streamer success, but on sustaining belief and dependence on paid services (\autoref{embedd}).

\subsection{Transforming Platform Speculation into Moral Labor Expectations} \label{trans}

As discussed in the previous \autoref{external}, MCNs interpret opaque and unstable platform algorithms by translating them into simplified rules and actionable advice. These interpretations serve specific organizational interests by creating the illusion that algorithmic success can be achieved through discipline, effort, and optimization. In this section, we show how such interpretations are systematically leveraged into labor management to shape how streamers work and evaluate themselves.

One major way this happens is through the transformation of speculative ideas about the algorithm into moral labor expectations. Beliefs such as “the algorithm rewards long hours” or “visual quality guarantees recommendations” are framed as certainties. These claims are repeatedly emphasized in on-boarding materials, group chat guidance, and informal coaching, where they gain authority as commonsense truths. Over time, they become internalized by streamers as standards of professional conduct.

Instead of merely helping streamers understand how the platform might work, MCNs use these beliefs to define how streamers should behave. The burden of visibility is shifted entirely onto the individual: if streamers are not being recommended, it is because they are not working hard enough, not investing enough on equipments, or not physically attractive or expressive enough, i.e., transforming the responsibility onto individual efforts. L6 recalled: \textit{“The MCNs I contacted told me I only needed to stream every day. Just show up, and the algorithm will recognize you. I thought, okay, then it’s just about effort.”} 

This framing naturalizes structural opacity as personal failure. M2 explained: \textit{“Most streamers think it’s their fault. That’s what keeps them going. They keep chasing that moment when the algorithm ‘notices’ them.”} Even when outcomes do not improve, many streamers continue to blame themselves. L7 described his own response: \textit{“They said if I streamed every day with more positive emotions expressions, I’d get algorithmic recommendations. So I smile even when I’m exhausted, because maybe today is the day the algorithm sees me.”}

Because streamers believe the system rewards the right behavior, they are willing to endure exhaustion, self-doubt, and financial risk. This makes the MCN’s job easier. As M11 put it: \textit{“We MCNs don’t need to spend much. If they streamers believe in the algorithm, they’ll work hard for free. No other incentives or strict management are needed. That’s the best-case scenario for any company.”}

Streamers act based on what they believe the algorithm expects, but those beliefs are largely shaped by the MCN’s interpretation. In L22’s case, the makeup choices were labeled as insufficient by MCNs not because of any clear platform rule, but because the MCN had already instilled the idea that visual performance is central to algorithmic success. As L22 put it, \textit{“They didn’t force me, but I felt like if I didn’t improve my makeup, I was being lazy or unprofessional.”}

These internalized expectations not only sustain streamers’ labor but also strategically align with the MCNs’ internal understanding of algorithmic uncertainty. Internally, MCNs know they cannot guarantee any single streamer’s success due to the volatility of recommendation algorithms. As such, offloading the responsibility of failure becomes an essential managerial tactic. It allows MCNs to avoid accountability for poor performance outcomes while maintaining the appearance of support and guidance.

\newadd{MCNs’ interpretation is central to labor control because many streamers remain in the industry precisely through this form of hope-driven labor. Without this horizon of hope, many streamers would exit the industry much earlier.} \newadd{F1 worked as a streamer for one year, first collaborating with an MCN for six months and then continuing independently for another six months. During her time with the MCN, she strongly believed that success was attainable and that the algorithm rewarded effort in a fair manner. This belief persisted even after conflicts with the MCN emerged. As a result, she continued live-streaming independently for an additional six months, holding onto the expectation that consistent labor would eventually pay off. It was only after sustained poor performance and declining income that she decided to leave the industry. F1 explained: \textit{“Many of us stay in this industry because we believe there is hope to make money. MCNs always describe other industries as less fair than live-streaming.”}}

\newadd{It is worth noting that most platforms, including Douyin, allow streamers to go live independently. Following the general principle, streamers do not need MCNs in order to participate in live-streaming. However, many streamers choose to join MCNs because they lack the necessary equipment, technical knowledge, and practical experience at the early stages of their careers. MCNs position themselves as providers of guidance, resources, and know-how, and their simplified narratives about algorithmic success make live-streaming appear worth sustained effort.} \newadd{These easily teachable interpretations are also central to MCNs’ ability to attract and retain a sufficiently large pool of streamers. Importantly, streamers are almost always free to leave. Especially after 2022, exit costs have become relatively low, with penalties described by MCNs as\textit{ “basically one month’s salary” }(M11). By framing success as something that can be achieved through effort, MCNs encourage streamers to join and to remain in the organization long enough to generate additional revenue streams, such as equipment rental fees discussed later in this paper. }

\newadd{Under these conditions, discursive construction becomes a key mechanism for labor control. Because streamers can exit at any time, control cannot rely on formal coercion. Instead, it depends on streamers’ willingness to stay, to keep investing labor, and to continue believing in the possibility of success.}

\subsection{Embedding Algorithmic Beliefs into Material and Contractual Structures} \label{embedd}

The interpretive work described in the previous section does not stop at motivation. MCNs extend their algorithmic explanations into the material, aesthetic, and contractual arrangements that define how streamers work. What begins as speculative beliefs, such as the idea that visual clarity or consistent appearance leads to higher algorithmic visibility, is gradually institutionalized into binding routines and financial commitments.

\newadd{Some equipment upgrades can indeed improve technical conditions, such as network stability or image clarity. However, MCNs actively amplify the significance of these improvements by framing them as decisive factors for algorithmic recognition and success. Through this framing, incremental technical enhancements are transformed into ongoing requirements for investment.} 

MCNs frequently tell streamers that clearer visuals and smoother audio will improve their ranking, reinforcing this belief through formal or informal expectations. L19 described his experience: \textit{“I spent 10{,}000 RMB on a Sony camera because the MCN said it would help me algorithmic visibility rank better. But nothing changed. They kept saying, maybe I need to improve my makeup or get a new background.”}

This expectation extends far beyond cameras. Ring lights, microphones, mixers, reflective backdrops, and LED panels are all framed as essential. In some cases, MCNs distribute internal “technical evaluation reports” that nudge streamers toward upgrades. M12 explained how this interpretation benefits the company: \textit{“Most streamers don’t buy equipment all at once. They rent from us first. If they start earning, they’ll invest more. If they don’t, at least we make money from the rentals.”} 

Alongside hardware, MCNs standardize streamers’ aesthetic presentation. Guidelines often dictate not only tone and posture, but makeup, camera angles, and even clothing schedules. M9 described a typical protocol: \textit{“They’re not allowed to wear the same outfit in a two-week cycle. We even advise which colors look better on camera. Some of them rent clothes from us.”} 

These requirements are enforced contractually through studio rules, performance evaluations, and surveillance systems. Even streamers who work from home are monitored. M4 explained: \textit{“We ask them to share their live link with a manager who checks if the lighting, camera, background are aligned with expectations. If not, they get a warning and we suggest them do a upgrade.”} 

\begin{figure}[htbp]
    \centering
    \includegraphics[width=\textwidth]{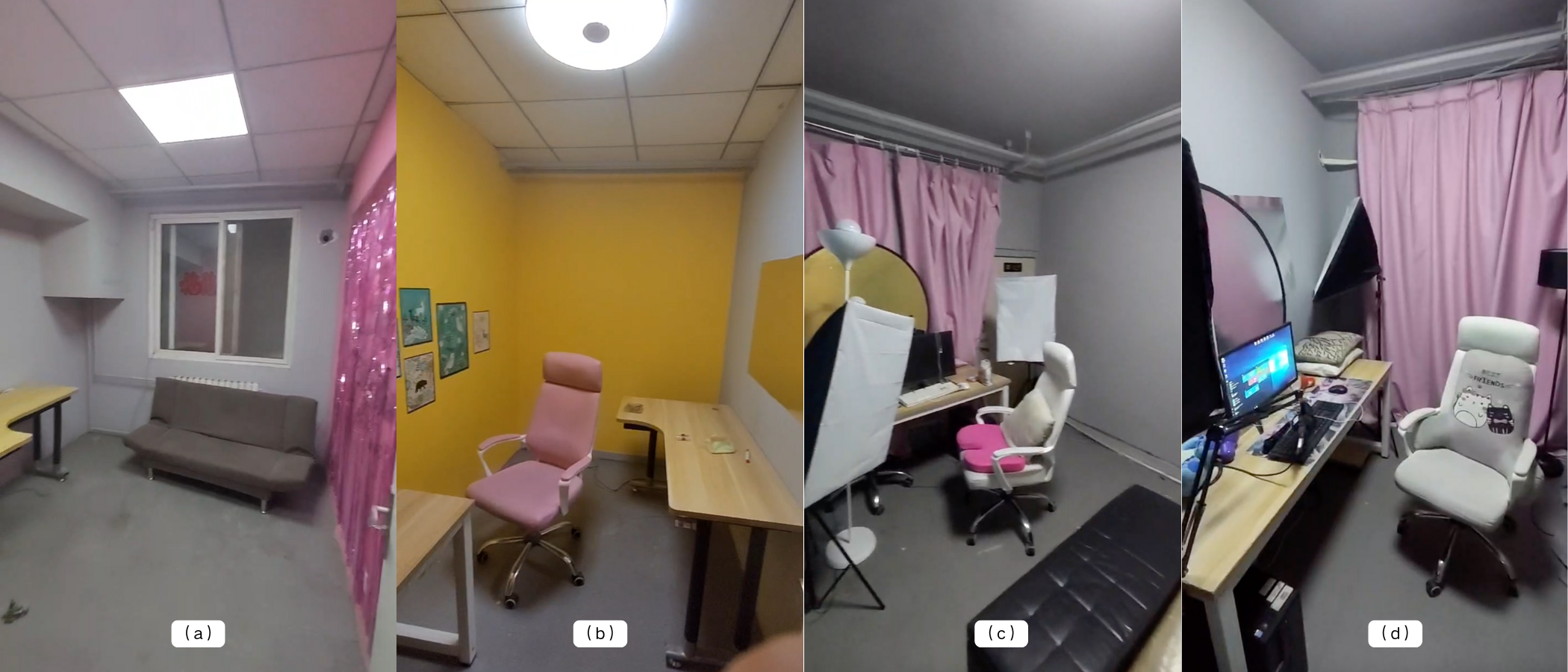}
    \caption{Comparative analysis of livestreaming studios with varying amenities and price points. All studios provide basic air conditioning, internet connectivity, and workspace furnishings, while differing substantially in equipment quality and decorative elements. (a) A bare basement dance streaming studio featuring a faux window facing a concrete wall, priced at approximately 2,500 RMB monthly without computer equipment, lighting systems, or professional backdrops beyond a rudimentary fringe curtain. (b) A compact basement studio designed for singing and conversational streaming, characterized by its confined windowless space, basic interior finishing, and pink swivel chair, available at approximately 2,200 RMB monthly. (c) A ground-floor multipurpose streaming studio equipped with fundamental lighting apparatus and computer systems, featuring a dedicated window, priced at approximately 3,100 RMB monthly. (d) A premium spacious studio situated on the second floor, furnished with an independent clothing rack, basic lighting equipment, professional microphone with stand, in-room audio monitoring speakers, dedicated window access, and a vanity mirror, available at 3,500 RMB monthly with the additional advantage of proximity to technical support staff offices for immediate assistance.}
    \label{fig:livestream_room_comparison}
\end{figure}

In some cases, MCNs also manage the physical layout of labor. Many operate shared studio spaces where streaming rooms are tiered by performance level. Higher-ranked streamers receive better infrastructure and services. L16 described how space and performance were linked: \textit{“At first, I shared a basement room with three other girls. The equipment frequently freezes or becomes unresponsive, it’s difficult to operate, and there’s poor air circulation. When I ranked higher one month, I moved to a private room upstairs. But I had to pay more for it, nearly 3{,}000 RMB a month.”} These layered expectations create cycles of escalating cost and performance pressure. L20 recalled: \textit{“I spent most of my early earnings on room rental, clothes, gear. I didn’t even realize I was in debt until month three. But by then, I didn’t dare quit; I was too invested.”} Streamers are often told that these are necessary “investments” in their future algorithmic visibility. But the platform algorithm offers no guarantees. L22 shared his experience: \textit{“I borrowed money from my cousin to redecorate my stream room. I told myself, if I looked more professional, I’d get better exposure. But nothing happened. Now I just feel stuck.”} As shown in ~\autoref{fig:livestream_room_comparison}, the material differentiation of studio spaces reflects a managed hierarchy of visibility and professionalism, reinforcing MCNs’ externally imposed belief that algorithmic success is contingent on aesthetic and technical upgrades.

As illustrated in \autoref{inter}, MCNs build streamers’ belief that algorithmic success depends on visible performance cues and technical polish. M12 captured the disciplining effect of this algorithmic interpretation: \textit{“We don’t force anyone to stay. But when you’ve rented equipment, decorated your room, signed the platform contract; it’s hard to walk away.”} 

\newadd{These behaviors and forms of self-understanding are shaped almost entirely by the algorithmic interpretations constructed and promoted by MCNs. In practice, live-streaming itself requires only minimal technical infrastructure, often no more than a smartphone with a built-in camera and microphone. Many platforms are basically functional under such basic conditions. The escalation of equipment purchases, studio upgrades, and aesthetic investments emerges only when streamers come to accept MCNs’ claims that algorithmic success depends on visible professionalism and constant upgrading. MCNs rarely compel streamers to make such purchases through direct coercion. Instead, investment is encouraged through interpretive frames that link spending to algorithmic recognition and future success.}

This behavior logic aligns with MCNs’ internal understanding of algorithmic unpredictability. Internally, MCNs acknowledge that algorithmic success cannot be reliably manufactured at the individual level. As such, they prefer to spread risk across a broad base of streamers, investing minimally while maximizing returns through volume and service monetization. The more streamers believe in the narrative that effort yields exposure, the more they will spend. In this way, the externally constructed algorithmic interpretations (that visibility is earned through discipline and polish) serves the internal interpretations (that success is probabilistic and unpredictable) by ensuring that MCNs profit from both success and failure, so long as belief is sustained.

\section{\newadd{Reactions to MCNs’ Discursive Construction of Algorithms: Live-Streamers’ Compliance, Doubt, and Departure (RQ3)}}\label{rq3}

\newadd{In the preceding sections, we examined how MCNs construct algorithmic interpretations (\autoref{RQ1}) and how these interpretations are mobilized as mechanisms of labor control (\autoref{RQ2}). It is important to clarify that MCNs’ capacity to govern live-streamers’ labor rests largely on algorithmic discourse and, crucially, on streamers’ acceptance of that discourse.} \newadd{On most major Chinese platforms, live-streamers are technically able to broadcast independently without affiliating with MCNs. Live-streaming itself requires only minimal equipment, and there is no inherent necessity to invest in professional devices or studio infrastructures. In practice, however, many streamers enter the live-streaming industry after encountering MCNs’ promotional narratives with optimistic career expectations. After joining, these expectations are continuously reinforced through everyday guidance and coaching. Carried by this promise of future success, large numbers of streamers enter the industry, collaborate with MCNs, continue to invest their labor, and willingly upgrade their equipment in pursuit of better streaming performance and increased algorithmic visibility.} 

\newadd{Because labor control is mediated through belief rather than coercion, streamers’ responses to MCNs’ algorithmic interpretations are not uniform. Some comply with these expectations and internalize them as professional norms (\autoref{reaction1}); others begin to question their validity as promised outcomes fail to materialize (\autoref{reaction2}); still others ultimately exit both MCNs and the live-streaming industry altogether (\autoref{reaction3}). In what follows, we analyze these divergent responses to show how algorithmic discourse is variably taken up, negotiated, and resisted in streamers’ everyday labor practices.}

\subsection{\newadd{Compliance with MCNs’ Labor Management}}\label{reaction1}

\newadd{Compliance is the most common response among live-streamers in our study. For many participants, accepting MCNs’ algorithmic interpretations represents the least psychologically and cognitively demanding option under conditions of algorithmic uncertainty. Rather than constantly questioning platform mechanisms or organizational guidance, compliance offers a sense of direction and emotional relief in an otherwise opaque and volatile environment.}

This tendency toward compliance is closely tied to the demographic composition of China’s live-streaming industry. The industry is heavily female-dominated and disproportionately attracts young women from rural areas, many of whom face constrained employment opportunities in other sectors of the labor market \cite{xiao2024happens}. For these workers, live-streaming often appears to be one of the few accessible pathways to flexible income generation or upward mobility. Within this context, MCNs’ narratives of algorithmic fairness, effort-based success, and eventual recognition become especially compelling.

For many streamers, compliance is therefore not a passive or naïve choice, but a pragmatic response to limited alternatives. As several participants emphasized, when everyday life is marked by economic precarity and uncertainty, any option that promises even a modest chance of improvement can feel worth pursuing. This pattern has been documented in prior research on China’s live-streaming industry \cite{xiao2024happens} and was also evident in our ethnographic fieldwork. Multiple participants (e.g., L1, L4, L6, L13, L16) noted that even if they left live-streaming, they had no clear sense of what other work they could realistically do.

This sense of constrained choice is also reflected in the demographic profiles of our participants (\autoref{tab:my_label1}), particularly in relation to their educational backgrounds and earnings. L6, who completed only high school, described her perceived options as extremely limited: \textit{“Working in an electronics factory is exhausting, but doing food delivery or driving for Didi requires renting a car, which is very expensive up front—and besides, both industries are dominated by men.”} In such circumstances, complying with MCNs’ guidance and continuing to invest effort in live-streaming can appear more reasonable than exiting the industry altogether.\footnote{\add{It is important to note that these findings do not contradict prior HCI research highlighting the empowering potential of live-streaming for marginalized groups \cite{tang2022dare,wang2024there}. Many participants in our study also described tangible improvements in their living conditions after becoming streamers. For instance, L6 explained that she was able to move from a crowded dormitory to a small studio apartment. Although both housing situations were modest by urban standards, this change still represented a meaningful improvement in her daily life. In this sense, exploitation and improvement coexist: live-streaming may expose workers to new forms of control while simultaneously offering incremental material benefits. Rather than opposing earlier accounts, our findings show how these dynamics are experienced together in streamers’ everyday lives. For some participants, the fact that live-streaming allowed them to live slightly better than before also made its exploitative aspects feel less severe. As L6 put it, \textit{“other jobs might be even worse.”}}}

\subsection{\newadd{Doubting MCNs’ Algorithmic Narratives}}\label{reaction2} 

\newadd{Doubt emerges when live-streamers begin to notice persistent discrepancies between MCNs’ algorithmic explanations and their own lived outcomes. As promised visibility fails to materialize despite sustained effort and investment, some streamers start to question whether the algorithm is as effort-sensitive as they were led to believe.}

\newadd{These doubts are often triggered by repeated experiences of stagnation or regression. Streamers described situations in which they followed MCNs’ guidance closely—maintaining long streaming hours, upgrading equipment, adjusting appearance, and optimizing interaction—yet observed little change in traffic or recommendations. Over time, such experiences erode confidence in the causal links promoted by MCNs’ narratives. As L9 put it, \textit{“I did everything they told me to do. I streamed longer, bought better lights, smiled more. But the numbers didn’t move. After a while, I started wondering if it really worked the way they said.”}}

\newadd{Doubt does not necessarily lead to immediate resistance or exit. For many streamers, questioning algorithmic explanations coexists with continued compliance. Streamers may privately suspect that MCNs’ guidance is overstated or selectively framed, while still following it in practice. This ambivalence reflects the high stakes of withdrawal: exiting the industry means giving up not only future income but also the possibility that success might still arrive. As L12 explained, \textit{“I knew something was off, but I kept going. If I stopped, then all the effort before would really be wasted.”}}

\newadd{Doubt is also shaped by asymmetric access to information. Streamers rarely have alternative explanations for poor performance beyond those provided by MCNs. Platform algorithms remain opaque, and peer comparison often reinforces uncertainty rather than clarity. When other streamers occasionally succeed, their success is frequently interpreted as confirmation that the system still works, rather than as evidence of randomness. This keeps doubt fragmented and individualized, rather than developing into collective critique.}

\newadd{As a result, doubt functions less as a direct challenge to MCNs’ authority and more as a fragile, unstable state. Streamers oscillate between belief and skepticism, hope and frustration, without a clear pathway toward resolution. This prolonged state of uncertainty allows MCNs’ algorithmic discourse to continue exerting influence, even as its credibility weakens. Doubt thus represents an intermediate response to algorithmic control—one that signals cracks in belief while still binding streamers to continued participation.}

\subsection{\newadd{Departure from the Live-Streaming Industry}}\label{reaction3}  

\newadd{Departure represents the point at which belief in MCNs’ algorithmic interpretations becomes untenable and continued participation can no longer be justified. Unlike compliance or doubt, departure involves a concrete break from both MCNs and, in many cases, the live-streaming industry itself. This decision typically follows prolonged periods of stagnation, accumulated financial pressure, and emotional exhaustion.}

\newadd{Notably, departure is rarely immediate. Several participants described an intermediate phase in which they disengaged from MCNs while continuing to stream independently. F1 and F7 both left their MCNs and attempted to live-stream on their own for a period before ultimately exiting the industry altogether. These transitions were motivated by a desire to test whether success was possible outside MCNs’ control and guidance. As F1 explained, \textit{“I wanted to see if it was really the MCN holding me back, or if I just needed more time. So I kept streaming on my own.”} When performance did not improve despite this shift, the promise of algorithmic fairness in live-streaming lost its credibility.}

\newadd{At the same time, independent streaming does not necessarily preclude success. F7, for example, accumulated approximately 28,000 followers after leaving his MCN and streaming independently, with most of this audience growth occurring outside the MCN system. His case illustrates that algorithmic visibility is not inherently dependent on MCN affiliation. A small number of independent streamers are able to develop their own understandings of platform dynamics and cultivate relatively stable audiences without organizational mediation.}

\newadd{However, such cases are exceptional rather than typical in China, given the generally low level of digital literacy among most Chinese streamers. Independent success requires not only time and sustained experimentation, but also the capacity to develop one’s own folk theories about algorithmic visibility, audience engagement, and platform rhythms. While independent success remains possible in principle, it remains far from accessible to most streamers in practice. As M1 explained, this disparity also helps account for the prominence of MCNs in China: \textit{“MCNs may be less prevalent in Western contexts because most streamers there have relatively high digital literacy, but that is not the case in China.”}}

\newadd{Other streamers attempted to resolve doubt by switching organizational affiliations. F5 moved between two different MCNs before leaving the industry, hoping that a change in resources or guidance might lead to different outcomes. This strategy reflects continued belief in the underlying algorithmic narrative: that success remains achievable under the right conditions. Only after repeated disappointment across organizational contexts did exit become a viable option.}

\newadd{These trajectories suggest that departure is not a rejection of live-streaming from the outset, but the outcome of sustained efforts to make the system work. Streamers often exhaust available strategies—compliance, optimization, organizational switching, and independent streaming—before concluding that further investment is unlikely to pay off. By the time they leave, many describe a sense of clarity mixed with resignation. As F7 put it, \textit{“At some point, I realized it wasn’t about working harder or finding the right MCN. I just couldn’t keep going like this.”}}

\newadd{We intentionally sampled streamers with different exit trajectories in order to capture this processual nature of departure. Departure marks the limit of MCNs’ discursive control: when hope, optimization, and even organizational switching no longer sustain participation, streamers disengage from the industry altogether.}

\newadd{To conclude, compliance, doubt, and departure reveal a contextual condition shaping MCNs’ discursive influence. Many streamers in our study have limited formal education and digital literacy, constraining their ability to independently interpret platform algorithms. Under such conditions, MCNs’ algorithmic explanations become the primary resources for making sense of uncertainty and performance outcomes.} \newadd{This constraint produces a form of constrained agency. Streamers voluntarily rely on MCNs’ narratives as a pragmatic strategy for navigating opaque platforms, while limited alternatives make such reliance difficult to avoid. Trust is sustained not through overt coercion, but through asymmetric access to algorithmic knowledge and interpretive authority.} \newadd{As long as MCNs’ explanations remain credible, they organize labor by shaping how effort, investment, and future success are understood. When credibility erodes, streamers move from compliance to doubt and ultimately to departure, marking the withdrawal of belief that underpins MCN-managed labor.}

\section{Discussion}

This study reveals how discursive construction of algorithms—informal beliefs about how algorithmic systems work—are actively constructed, formalized, and exploited by intermediary institutions. Drawing on nine months of ethnographic fieldwork in China’s live-streaming industry, we show how Multi-Channel Networks (MCNs) strategically develop dual interpretations about algorithms: internally, they treat algorithms as opaque and unpredictable; externally, they frame algorithms as learnable and responsive. These interpretations do not simply interpret the algorithm—they organize labor around it.

\newadd{Our ethnographic findings illuminate the organizational dynamics that underpin these opportunities. Many MCNs operate under sustained pressure to recruit new streamers and to keep existing streamers active. In practice, MCNs often take a proactive stance toward streamers, investing considerable effort in outreach, onboarding, and persuasion. Streamers described signing with MCNs as a relatively flexible and low-commitment decision: many felt able to choose among agencies they preferred, and exit costs were generally low, both contractually and financially. Continued streamer participation nonetheless remains a central condition for MCNs’ own survival and profitability.} \newadd{This organizational dependence shapes how MCNs engage with streamers over time. Rather than relying primarily on coercive control, MCNs engage in ongoing discursive work to sustain motivation and align streamer expectations with organizational needs. Algorithmic interpretations play a central role in this process.}

\newadd{Concretely, MCNs circulate an external-facing algorithmic narrative built around two claims: (1) the algorithm is meritocratic and rewards persistence; (2) visibility follows identifiable signals that can be optimized;  (\textit{\textbf{RQ1}}). They translate the promised signals into mandatory-like routines (fixed schedules, performance targets, appearance standards, and continuous affective display) and then tie compliance to material arrangements that generate revenue for MCNs. Profit comes from commissions on successful streamers’ earnings, plus more reliable income from services sold to a large pool of newcomers, including training packages, equipment rental or sales, studio fees, and platform-operation support. The same narrative that attributes success to optimization also legitimizes recurring purchases and escalating investment (\textbf{\textit{RQ2}}). Streamers’ responses cluster into three trajectories. Many comply because the narrative offers direction and reduces cognitive burden under opacity; compliance also feels low-risk given flexible contracts and low exit penalties. Others express doubt when repeated effort and spending fail to change traffic, yet doubt often coexists with continued compliance due to sunk costs and the hope of a turnaround. Departure occurs when low returns, debt, and exhaustion accumulate to the point where the narrative no longer sustains continued participation, sometimes after a short phase of independent streaming to test whether MCN mediation was the bottleneck (\textit{\textbf{RQ3}}).}

Our findings suggest that MCNs’ discursive construction of algorithms functions not merely as interpretive framework, but as organizational infrastructures through which labor is managed. While much of the CSCW literature has emphasized algorithmic interpretation as emergent, user-driven responses to algorithmic opacity (e.g., folk theories) ~\cite{shen2021everyday,li2023participation,deng2023understanding}, our study shifts the analytic focus to institutional actors and their role in intentionally constructing and instrumentalizing such algorithmic interpretations. To develop this argument, the discussion proceeds in three parts. First, we pay attention to the organizational discursive construction of algorithms in the platform economy (\autoref{from}). Second, we analyze how these discursive construction of algorithms serve as instruments of labor governance, enabling MCNs to manage precarity, displace responsibility, and stabilize labor output through soft control (\autoref{labor}). Third, we reflect on the broader design and policy implications of these findings. We call for systems and policies that support epistemic transparency, making visible not just what algorithms do, but how knowledge about them is constructed (\autoref{implications}).

\subsection{\newadd{Constructing Algorithmic Authority: MCNs’ Discursive Construction of Algorithms}}\label{from}

\newadd{A key contribution of this study is to shift attention from user-centered interpretations of algorithms to the organizational production of algorithmic meaning. Prior scholarship on algorithmic interpretations, including work on folk theories, has largely examined how users or communities develop informal, bottom-up understandings to make sense of opaque algorithmic systems \cite{bucher2018if, devito2017algorithms, eslami2016first, xiao2024let}. These interpretations are typically described as experiential and emergent, shaped through everyday interaction, peer exchange, and speculation. Within this framing, algorithmic interpretations function as practical sensemaking resources that help users cope with uncertainty and informational asymmetries.}

\newadd{In contrast, our study focuses on how algorithmic interpretations are discursively constructed by organizations. In the context of MCNs, algorithmic knowledge does not primarily arise from individual trial and error, but is strategically produced, stabilized, and circulated through organizational discourse. These interpretations are embedded in routine training, guidance, and evaluation practices, forming part of the epistemic infrastructure through which MCNs manage and coordinate streamer labor. This shift in analytic focus highlights that algorithmic knowledge is not merely incidental or interpretive, but infrastructural. MCNs act as epistemic intermediaries that claim authority over how algorithms should be understood and acted upon, translating platform opacity into actionable narratives aligned with organizational goals. As such, the discursive construction of algorithms operates as a form of working infrastructure \cite{pipek2009infrastructuring, star1996steps, karasti2018studying, ludwig2018designing} that shapes labor practices, expectations, and investment, rather than simply reflecting users’ interpretations.}

\newadd{Importantly, the epistemic authority of MCNs does not come from possessing “correct” or privileged knowledge about platform algorithms. Their understandings are partial, heuristic, and contingent. MCN staff themselves routinely acknowledged that their recommendations may work in some cases yet fail in others, and that they cannot predict outcomes for any specific streamer with confidence. MCNs’ power instead comes from their ability to author and institutionalize algorithmic interpretations under conditions of opacity. In this sense, labor control is generated less by “knowing the algorithm” and more by controlling the narratives through which algorithmic outcomes are explained, attributed, and acted upon. Without the infrastructuralization of algorithmic explanations, MCNs would struggle to sustain their organizational operations, lacking a credible basis for recruitment, retention, and revenue generation, rendering their intermediary role difficult to maintain.} \newadd{While existing scholarship on folk theories has richly documented grassroots sensemaking under algorithmic opacity, it has rarely examined how algorithmic explanations become institutionalized as durable cognitive infrastructures. Our findings foreground this organizational transformation of algorithmic interpretation and its consequences for labor governance.}

Second, our findings challenge the prevailing binary in CSCW and HCI that separates algorithmic knowledge into formal (e.g., platform documentation, API guides) and informal (e.g., user speculation, peer discussion, online rumor) categories~\cite{kusk2022platform,xiao2024let,bishop2019managing}. In the context of MCNs, this division collapses. The algorithmic knowledge that circulates within these organizations is neither officially sanctioned nor casually improvised. Instead, we conceptualize it as MCNs’ discursive construction of algorithms—informal knowledge that is produced and stabilized by intermediary actors who lack direct access to platform-engineered formal knowledge. MCNs are not platform insiders; they operate without formal connections to algorithm developers or privileged access to system logics. Yet they construct and disseminate their own working interpretations of “how the algorithm works” and embed these into onboarding scripts, coaching sessions, ranking systems, and contractual expectations. These discursive constructions gain a quasi-formal status—not through technical verification, but through routinized organizational practice, repetition, and discursive reinforcement. In doing so, MCNs transform folk interpretation into a mechanism of labor governance. 

\newadd{Third, a key feature of these discursive constructions lies in the tension between MCNs’ organizational practices and their public claims. As companies, MCNs operate under conditions of limited control over algorithmic outcomes. Internally, they manage risk through batch recruitment, uneven resource allocation, and service-based monetization that does not depend on any individual streamer’s success. Externally, however, MCNs present themselves as capable algorithmic guides. This contrast allows MCNs to reconcile uncertainty with authority: organizational practices are oriented toward probabilistic returns and volume-based profit, while external discursive claims foreground individualized improvement and attainable success for steamers. The gap between what MCNs do as firms and what they claim about the algorithm is not incidental. It is precisely this gap that enables discursive constructions to function as labor-governing infrastructure, sustaining participation and legitimizing continued investment despite unstable outcomes.}

\newadd{Moreover, we argued that the belief in MCNs’ algorithmic interpretations reflects unequal capacities for knowledge production within platformized labor markets. In contrast to MCNs’ discursive construction of algorithms, prior research shows that experienced content creators can develop their own informal understandings of platform behavior through prolonged engagement and experimentation \cite{devito2022transfeminine,wu2019agent,moran2022folk}. Our ethnographic data indicate that most streamers in MCN-managed live-streaming ecosystems lack the material and epistemic conditions required to develop comparable independent interpretations. Many participants entered live-streaming with limited prior experience in content production, unstable offline employment, and strong pressure to generate income quickly. Under these conditions, systematic experimentation with content strategies or algorithmic responses was difficult to sustain. MCNs occupy this gap by offering ready-made explanations of how the algorithm works and how effort should be organized. These explanations are persuasive not because they are demonstrably accurate, but because they provide a coherent account that reduces uncertainty and distributes responsibility. For streamers who lack the resources to independently test or verify algorithmic behavior, trusting MCNs’ interpretations becomes a rational response to structural constraints rather than a simple failure of critical judgment.}

\subsection{Discursive Construction of Algorithms for Labor Management} \label{labor}

While most scholarship on algorithmic interpretations emphasized their adaptive or empowering qualities, helping users make sense of opaque systems in the absence of formal knowledge \cite{huang2022folk,siles2020folk,bonini2024algorithms,eslami2016first}, our findings suggest a more complex and strategic function: algorithmic interpretations are not only interpretive frameworks, but also labor governance tools. In the context of live-streaming platforms, they become a key mechanism through which institutions like MCNs direct labor, manage uncertainty, and externalize risk. 

A growing body of CSCW and HCI literature has investigated the dynamics of algorithmic management, especially in gig economies and platform labor systems~\cite{kellogg2020algorithms,mohlmann2021algorithmic}. Here, scholars have shown how algorithms allocate work, evaluate performance, and discipline workers without the need for traditional human supervision. However, this line of work often focuses on direct algorithmic feedback (e.g., ratings, scores, rankings) or automated decision-making (e.g., job assignments). Our findings build on this by revealing a second layer of algorithmic management: the interpretive layer, where algorithmic outputs are not simply accepted or resisted, but actively narrated, framed, and naturalized through algorithmic interpretations constructed by intermediaries. These algorithmic interpretations are not mere byproducts of algorithmic opacity but are engineered and circulated by MCNs to align with institutional interests and govern labor at scale.

\newadd{Central to this mode of labor governance is a pronounced epistemic asymmetry between MCNs and live-streamers. MCNs occupy a privileged position in the circulation of algorithmic knowledge: they control when, how, and in what form explanations about algorithmic behavior are made available to streamers. Although MCNs do not possess verified or complete knowledge of platform algorithms, they nevertheless monopolize interpretive authority by presenting themselves as the most credible source of explanation in an otherwise opaque system. Streamers, by contrast, typically lack the resources, comparative data, and temporal distance needed to independently evaluate algorithmic claims or to develop alternative explanatory frameworks.}

\newadd{This asymmetry is not merely informational but organizationally produced. MCNs’ internal recognition of algorithmic uncertainty remains largely inaccessible to streamers, while externally circulated interpretations are selectively simplified, stabilized, and framed as actionable truths. As a result, streamers encounter algorithmic outcomes through a narrowed interpretive channel, where uncertainty is individualized and responsibility is localized. Even when outcomes fail to materialize, streamers often lack the epistemic grounds to contest MCNs’ explanations, leading doubt to remain fragmented and personalized rather than collective or structural.}

\newadd{Through this asymmetry, MCNs are able to convert limited and speculative knowledge into durable authority. Control is exercised not by withholding information entirely, but by curating the forms of knowledge that circulate and defining what counts as legitimate understanding of algorithmic success and failure. In this sense, epistemic asymmetry becomes a key condition for the effectiveness of discursive construction: it allows algorithmic interpretations to function as governance tools precisely because alternative interpretations are difficult to verify, share, or stabilize among workers.}

MCNs therefore construct what we call interpretive infrastructures—a system of informal yet durable knowledge that governs through belief, not verification/reality. Our findings contribute to CSCW’s growing attention to the epistemic and organizational dimensions of algorithmic labor. Algorithmic interpretations, we argue, are not peripheral but are central to how algorithmic management operates in opaque, high-risk environments. They enable institutions to govern through the circulation of strategically engineered narratives. This invites future research into how informal knowledge can become a durable, institutionally mediated mode of control in algorithmic workplaces.

\add{Additionally, a deeper exploration of the dynamic between external and internal algorithmic interpretations would also enrich this discussion around labor management. In our study, external algorithmic interpretations constructed by MCNs serve managerial interests, stabilizing labor through prescriptive narratives. In contrast, recent work highlights how individual users develop their own folk theories as diagnostic tools for navigating platform logics and adapting their own practices \cite{mayworm2024content}. This comparison underscores two divergent trajectories: organizational actors such as MCNs strategically engineer algorithmic interpretations to manage others’ labor, while users mobilize folk theories to make opaque systems actionable for themselves. Future inquiries must therefore pay closer attention to whose interests these algorithmic interpretations represent and what purposes they serve. Importantly, algorithmic interpretations should not be assumed as inherently empowering or exploitative; rather, they occupy an ambivalent position that can be instrumentalized in different directions. For labor researchers, unions, and advocates, this means recognizing that algorithmic interpretations are 
contingent tools whose effects depend on the power relations and institutional agendas through which they are mobilized.}

\subsection{Design and Policy Implications} \label{implications}
Our findings suggest that designing for algorithmic systems in collaborative environments requires a shift in focus, from revealing how algorithms technically function to illuminating how algorithmic meaning is constructed, circulated, and institutionalized. In practice, workers interact not with raw code or backend logic, but with interpretations of the algorithm, often delivered and enforced through intermediary actors like MCNs or online communities ~\cite{rosenblat2016algorithmic,bishop2020algorithmic}. Especially under conditions of opacity, these interpretations acquire operational power, shaping behavior and structuring labor.

Thus, design efforts should support what we call epistemic transparency: making visible not just the algorithm’s outcomes, but the infrastructures that produce and validate its meaning. Our study shows that such infrastructures are often embedded in mundane but powerful organizational practices. For instance, MCNs disseminate algorithmic “best practices” through scripted onboarding sessions and coaching routines, which not only instruct streamers on how to behave but also encode speculative beliefs about what the algorithm rewards. These beliefs are further materialized through equipment sales, where streamers are encouraged to purchase lighting, cameras, and décor upgrades as investments in algorithmic visibility. Rather than merely functional, these practices reflect and reinforce particular interpretations of the algorithm that benefit intermediary institutions while externalizing risk and responsibility onto workers. Designing for epistemic transparency thus requires surfacing such hidden infrastructures and the interests they serve, while also enabling alternative interpretations, such as folk theories emerging from peer networks, to support more pluralistic and contested understandings of algorithmic systems. 

We also highlight epistemic asymmetry ~\cite{ajmani2024whose,ajmani2023epistemic} as a critical design and policy concern in algorithmic labor. Control over what is knowable about algorithms can be as powerful as control over the algorithm itself. To counter this, future systems might allow workers to annotate, challenge, or collaboratively construct algorithmic explanations, visualize competing interpretations, or track how dominant narratives shift over time. From a policy perspective, our findings call for stronger accountability mechanisms around algorithmic sensemaking. Platforms should not only disclose changes in algorithmic logic but also audit how such changes are communicated and operationalized by third parties such as MCNs. Regulatory frameworks might require platforms to monitor and report how algorithmic knowledge is intermediated, particularly when it becomes a basis for labor discipline or financial extraction. Policies could also mandate the provision of accessible explanatory resources, designed with input from workers and labor advocates, to mitigate dependency on opaque, potentially exploitative coaching ecosystems.

Finally, our work invites CSCW researchers, designers, and policymakers to view algorithmic interpretations as an organizational labor management process. Institutions do not merely allocate labor or payment, and mediate access to meaning. Rather than only designing tools to increase algorithmic transparency, we invite CSCW and HCI scholars to ask: who gets to construct algorithmic meaning, for whom, and to what ends? Designing technologies for algorithmic collaboration therefore requires grappling with the political economy of epistemic authority and these questions raise broader design and policy challenges for shaping the future of algorithmic work.

\section{Limitations and Future Work}
As with all qualitative research, our findings are shaped by the specific contexts, participants, and interpretive lenses of our study. There are a few important limitations. First, this research draws on ethnographic fieldwork conducted in three MCNs in China, and while we prioritized diversity in location, streamer profile, and organizational structure, our sample cannot represent the full heterogeneity of the live-streaming industry. Second, our focus on China brings unique sociopolitical and economic conditions that influence how platform labor and algorithmic governance unfold. While these conditions offer valuable insights into the global dynamics of platform intermediation, they may not be directly generalizable to other contexts. Third, our access to internal MCN practices was inevitably limited: despite prolonged immersion and trust-building, some dimensions of organizational decision-making and platform negotiation remained opaque. \newadd{Relatedly, our study foregrounds epistemic asymmetry but does not systematically examine how such asymmetries vary across organizational settings or how they might be mitigated in practice. Future research could investigate the conditions under which interpretive authority over algorithms is concentrated or redistributed, as well as the implications of these dynamics for workers’ agency.} Finally, our analysis focuses on the discursive construction of algorithms, but does not directly evaluate the technical validity of algorithmic interpretations themselves or track their longitudinal impact on streamer outcomes—areas that future mixed-method studies could explore.

\section{Conclusion}

This study examined how algorithmic labor is governed through the interpretive and organizational practices of intermediary institutions besides platforms themselves. Rather than focusing only on platform power or individual behaviors, our analysis highlights how Multi-Channel Networks (MCNs) in Chinese live-streaming industry play a central role in constructing algorithmic interpretations. We identify a system of dual algorithmic interpretations: internally, MCNs acknowledge the volatility and opacity of platform systems, adopting probabilistic strategies to manage organizational risk; externally, they promote simplified, prescriptive narratives that portray the algorithm as transparent, fair, and responsive to individual effort. We argue that these interpretations, though speculative, unverifiable, and often inaccurate, can be mobilized as tools of labor control: providing streamers with simplified interpretations of algorithmic success, converting platform uncertainty into actionable scripts, and shifting responsibility for failure onto individuals. We call for future research to focus on the role of intermediary institutions in algorithmic management and to attend to how discursive construction of algorithms  shapes the possibilities and limits of platform labor. By tracing how algorithmic interpretations are used not only to explain uncertain algorithms, but to manage platform labor, this study reveals how interpretation itself becomes a key site of power in the algorithmic age.

\bibliographystyle{ACM-Reference-Format}
\bibliography{sample-base}

\appendix

\end{document}